\documentclass[twocolumn,times]{aastex63}

\usepackage{dsfont}

\shorttitle{Observational signature of planet migration}
\shortauthors{Weber et al.}

\newcommand{\rhog}{\rho_{\rm g}}

\newcommand{\OmegaK}{\Omega_{\rm K}}

\newcommand{\dr}{\delta\rho}
\newcommand{\de}{\delta}

\newcommand\aip{Leibniz-Institut f{\"u}r Astrophysik Potsdam (AIP),
	An der Sternwarte 16, 14482, Potsdam, Germany}
\newcommand\nbia{Niels Bohr International Academy, The Niels Bohr Institute, University of Copenhagen,
	Blegdamsvej 17, DK-2100 Copenhagen {\O}, Denmark}

\begin{document}
\title{Predicting the observational signature of migrating Neptune-sized planets in low-viscosity disks}

\author{Philipp Weber}
\affiliation{\nbia}

\author[0000-0003-2953-755X]{Sebasti\'an P\'erez}
\affiliation{Universidad de Santiago de Chile, Av. Libertador Bernardo O'Higgins 3363, Estaci\'on Central, Santiago, Chile}
\affiliation{Departamento de F\'isica, Universidad de Santiago de Chile, Av. Ecuador 3493, Estaci\'on Central, Santiago}

\author{Pablo Ben{\'i}tez-Llambay}
\affiliation{\nbia}

\author[0000-0002-5398-9225]{Oliver Gressel }
\affiliation{\aip}\affiliation{\nbia}

\author[0000-0002-0433-9840]{Simon Casassus}
\affiliation{Departamento de Astronom\'ia, Universidad de Chile, Casilla 36-D, Santiago}

\author[0000-0001-7671-9992]{Leonardo Krapp}
\affiliation{\nbia}

\email{$^\dagger$Philipp.Weber@nbi.ku.dk}
\date{\today}

\begin{abstract}
  The migration of planetary cores embedded in a protoplanetary disk is an important mechanism within planet-formation theory, relevant for the architecture of planetary systems. 
  Consequently, planet migration is actively discussed, yet often results of independent theoretical or numerical studies are unconstrained due to the lack of observational diagnostics designed in light of planet migration.
  In this work we follow the idea of inferring the migration behavior of embedded  planets by means of the characteristic radial structures that they imprint in the disk's dust density distribution. We run hydrodynamical multifluid simulations of gas and several dust species in a locally isothermal $\alpha$-disk in the low-viscosity regime ($\alpha=10^{-5}$) and investigate the obtained dust structures. In this framework, a planet of roughly Neptune mass can create three (or more) rings in which dust accumulates.  
   We find that the relative spacing of these rings depends on the planet's migration speed and direction. By performing subsequent radiative transfer calculations and image synthesis we show that -- always under the condition of a near-inviscid disk -- different migration scenarios are, in principle, distinguishable by long-baseline, state-of-the-art ALMA observations.
\end{abstract}

\keywords{planetary systems: protoplanetary disks --
					hydrodynamics --
					radiative transfer --
					planet-disk interactions --
					submillimeter 
               }

\section{Introduction}
Observations of protoplanetary disks (PPDs) in recent years have revealed impressive substructures in the dust distribution of many targets \citep[e.g.][]{2015ApJ...808L...3A,2016ApJ...820L..40A,2018ApJ...869L..41A}. A large fraction of the observed disks show annular intensity enhancements or deficits, commonly described as rings and gaps, respectively. These remarkable concentric features can be explained by the radial accumulation and depletion of solid material at these locations and have elicited a number of proposed physical explanations. For example, the effect of magnetic fields seems to be significant in both the ideal and the non-ideal regimes of magneto-hydrodynamics \citep[MHD,][]{2015A&A...574A..68F,2016A&A...590A..17R,2018ApJ...865..105K,2019A&A...625A.108R}. Also the density and dust opacity modification expected at ice lines is promising for systems, where the radial position of the observed gaps coincides with expected phase transition temperatures \citep{2015ApJ...806L...7Z,2015ApJ...815L..15B} or where grain size evolution by sintering becomes efficient \citep{2016ApJ...821...82O}.

Moreover, the presence of a planet is a popular explanation for gaps and rings. If the mass of an embedded body exceeds a certain threshold, the gravitational interaction with its environment creates an underdensity along its orbit and consequently carves a gap. This was first shown for the impact of giant planets on gas dynamics \citep[e.g.][]{1986laP} but has been extended to study the behavior of dust in disks hosting giant planets \citep{2006A&A...453.1129P,2006MNRAS.373.1619R} or planets as small as a few Earth masses \citep[e.g.][]{2016MNRAS.459.2790R}. Decisive for the outcome, besides the mass of the embedded planet, is the level of turbulent viscosity present in the protoplanetary environment. For many years, the level of viscosity has been estimated from stellar accretion rates, associating it with the mechanism of outward transport of angular momentum and hence regarding viscosity as the driver of the stellar accretion process. 
However, it has been shown that the magneto-rotational instability \citep[MRI,][]{1991ApJ...376..214B} -- once believed to generate turbulence and from this the required level of viscosity -- is not fully active when considering the non-ideal MHD effects for typical conditions of a PPD and instead magnetic winds are expected to be the main carrier of angular momentum \citep{2013ApJ...769...76B,2015ApJ...801...84G}.
These numerical results, along with the observational upper limit estimates for turbulence -- found from line observations \citep{2017ApJ...843..150F,2018ApJ...864..133T} and the vertical extent of observed dust structures \citep{2016ApJ...816...25P} -- suggest, that the level of turbulent viscosity is considerably lower than what is needed to explain observed stellar accretion rates purely through viscous angular momentum transport. On the other hand, it is quite plausible that the disk is not completely inviscid, but a certain level of turbulence is still retained, generated by purely hydrodynamical processes, such as the vertical shear instability \citep[VSI,][]{2013MNRAS.435.2610N,2014AA...572A..77S}. The effect of a planetary perturbation on the density structure has been studied for inviscid disks using linear theory \citep{2001ApJ...552..793G} and also for disks of low viscosity by performing hydrodynamical simulations \citep[e.g.][]{2017ApJ...843..127D,2017ApJ...850..201B}. The latter studies suggest that a single planet embedded in the disk causes a series of concentric rings to form, both inside and outside of the planet's orbital location. 

The efficiency of many proposed processes at different stages of planet formation depends on the radial distance to the star at which one assumes the planet to form. The catalogue of detected exoplanets only offers valuable information to the question where planets typically end up. It does not necessarily tell us where those planets are forming.
This is due to the mechanism of planet migration, a process which commonly describes the modification of the planet's semi-major axis due to interaction with its environment. Migration is therefore a key parameter for the studies of population synthesis \citep[as reviewed in][]{2014prpl.conf..691B} that are engaged in linking the exoplanet data to early stage processes. 

The theory of planet migration is an extensive subject. It has been shown that the disk exerts a torque onto the planet, i.e. it changes the bodies angular momentum and causes it to move radially. For comprehensive reviews of the classical picture we refer to \citet{2012ARA&A..50..211K} and \citet{2014prpl.conf..667B}. Typically, this movement is directed towards the central star \citep{2002ApJ...565.1257T,2003ApJ...588..494M}. Some recent work highlighted, however, that an actively accreting embedded planet experiences an additional torque through its heating \citep{2015Natur.520...63B} and through the asymmetric feature in the dust distribution of the planetary gap \citep{2018ApJ...855L..28B} which are typically directed outwards and might even reverse the direction of migration. The exact process in a low-viscosity disk is challenging to describe and to study \citep{2019MNRAS.484..728M}, both from an analytical or numerical point of view, since the migration rate strongly depends on the assumed disk model and the processes taken into account. Many key parameters of PPDs cannot be constrained by direct observation such that migration rates obtained from theoretical models are often not directly applicable.
 Nevertheless, the gap structure of a migrating planet can change substantially compared to a planet on a fixed orbit, as shown by \citet{2019MNRAS.482.3678M} in the case of a high level of disk viscosity, an effect that was tested for observability in \citet{2019MNRAS.485.5914N}. 
 
 Only recently, \citet{2019AJ....158...15P} observed an interesting structure in the specific example of the disk around HD~169142, which they find to be inconsistent with a planet on a fixed orbit. Instead, \citet{2019AJ....158...15P} are able to explain the detected features with an inwards migrating planet in a poorly viscous environment.
 
 While these works rely on their specific model of simulated planet migration to be adapted to the specific system at hand, we utilize a bottom-up approach to obtain more clarity about the impact of planet migration in general.  
 For this we vary the migration rate by a simple prescription and monitor how the structure of the PPD reacts to each different scenario. 
 This idea has been previously followed by \citet{2017ApJ...843..127D} for two migration rates and considering only one dust species. Here, we expand this approach to include multiple dust species and a larger set of migration rates. While in the works of \citet{2019MNRAS.482.3678M} and \citet{2019MNRAS.485.5914N} the disk's viscosity has a crucial impact on its response to the planet's potential, we here consider PPDs that are
 nearly inviscid. In this scenario the waves launched by the planet are less damped and sharper structures can develop. Our intent is to characterize the most robust features a certain migration rate produces in a certain PPD and to link these properties to observable signatures in state-of-the-art observations.\\
The paper is structured as follows: In Section~\ref{sec:disk-model} we describe the employed disk model, as well as the setup for the fiducial hydrodynamical simulations, results of which are presented in Section~\ref{sec:hydro}. In Section~\ref{sec:synt-observ} we perform radiative transfer on the hydrodynamical results, followed by image synthesis to assess different migration scenarios for observability in interferometric observations. Section~\ref{sec:parameters} extends the hydrodynamical simulations to highlight the effects of important parameters and in Section~\ref{sec:discussion} we discuss caveats to our results and possible applications.

\section{Disk Model}\label{sec:disk-model}
\subsection{Governing Equations}
This work studies the coupled temporal development of the gas and several dust species, differing in the grain size they represent, in presence of a migrating planet. Both gas and dust are treated as fluids and their spatial distribution is described by their surface densities, $\Sigma_\mathrm{g}$ and $\Sigma_{i}$, respectively. The subscript $i$ distinguishes between different dust species and iterates from $i=1$ to $i=\mathrm{N}_\mathrm{dust}$, the number of dust species. The evolution of the considered fluids is described by the continuity and Navier-Stokes equations:
\begin{eqnarray}
&&\frac{\partial\Sigma_\mathrm{g}}{\partial t} + \mathbf{\nabla}\cdot \left(\Sigma_\mathrm{g}\mathbf{u} \right)   \, = \, 0 \,, \label{eq:contgas}  \\[5pt]
&&\Sigma_\mathrm{g}\left( \frac{\partial  \mathbf{u}}{\partial t} + \mathbf{u}\cdot \mathbf{\nabla}  \mathbf{u}\right)
\, = \,   - \mathbf{\nabla}P - \mathbf{\nabla}\!\cdot\!\tau \; - \Sigma_\mathrm{g}\mathbf{\nabla \phi}  -  \sum_{i}\Sigma_{{i}} \mathbf{f}_{{i}}\,.
\label{eq:NS-gas}  \\[5pt]
&&\frac{\partial\Sigma_{{i}}}{\partial t} + \mathbf{\nabla}\cdot \left(\Sigma_{{i}}\mathbf{v}_{{i}} +\mathbf{j}_{{i}}\right)  =  0 \,, \label{eq:contdust}\\[5pt]
&&\Sigma_{{i}} \left( \frac{\partial  \mathbf{v}_{{i}}}{\partial t} + \mathbf{v}_{{i}}\cdot \mathbf{\nabla}  \mathbf{v}_{{i}}\right)\;  =  \; -\Sigma_{{i}}\mathbf{\nabla \phi} + \Sigma_{{i}} \mathbf{f}_{{i}}\,.
\label{eq:NS-dust}
\end{eqnarray}
The velocity vectors of gas and dust species are denoted by $\mathbf{u}$ and $\mathbf{v}_{{i}}$, respectively. The isothermal gas pressure is defined as $P=\Sigma_{\rm g} c_\mathrm{s}^2$, with $c_\mathrm{s}$ being the speed of sound. The dust fluid is assumed to be pressureless. The viscous stress tensor is given by 
\begin{equation}
\tau \equiv \Sigma_{\rm g}\, \nu \left[\; \mathbf{\nabla} \mathbf{u} + (\mathbf{\nabla}\mathbf{u})^T - \frac{2}{3}(\mathbf{\nabla}\cdot \mathbf{u})\,\mathds{1}\;\right]\,,
\end{equation}
where $\nu$ is the kinematic viscosity and $\mathds{1}$ the identity matrix. To model viscosity in the disk, we employ the $\alpha$-viscosity framework described in \citet{1973A&A....24..337S}, where the viscosity is coupled to the thermal disk structure by a dimensionless constant $\alpha$:
\begin{equation}
\nu = \alpha c_\mathrm{s} h r\,.
\end{equation}
The aspect ratio, $h=Hr^{-1}$, compares the disk's pressure scale-height, $H=c_\mathrm{s}/\Omega_\mathrm{K}$ (with $\Omega_{\rm K}$ being the Keplerian angular frequency) to the radial extent of the disk. 
The combined gravitational potential of the star and the planet that appears in Equation~\ref{eq:NS-gas} and Equation~\ref{eq:NS-dust} reads
\begin{equation}
\phi = -\frac{Gm_\ast}{r} -\frac{G\,m_{\mathrm{p}}}{\left(|\mathbf{r}-\mathbf{r}_{\mathrm{p}}|^2+(bH_\mathrm{p})^2\right)^{\frac{1}{2}}} + \frac{G\,m_{\mathrm{p}}}{r_{\mathrm{p}}^2}r\cos{\varphi}\,,
\label{eq:gravpot}
\end{equation}
with the gravitational constant, $G$, the stellar mass, $m_\ast$, the planet mass, $m_\mathrm{p}$, and its position relative to the star, $\mathbf{r}_\mathrm{p}$. The parameter $b$ represents the smoothing factor which multiplied by the scale height at the planet's location, $H_\mathrm{p} = H(r_\mathrm{p})$, accounts for the vertical extent of the gravitational potential. In our models its value is set to $b=0.6$ \citep{Masset2002,Mueller2012}. The last term in Equation~\ref{eq:gravpot} corresponds to the non-inertial acceleration in the employed astrocentric frame of reference.

The quantity $\mathbf{f}_{i}$ in Equations~(\ref{eq:NS-gas}) and (\ref{eq:NS-dust}) represents the collisional interaction between gas and dust. In the regime that is relevant for this work (the so-called Epstein regime), the strength of this force depends linearly on the velocity difference between gas and dust fluids \citep{Safronov1972,1972fpp..conf..211W}: 
\begin{equation}
\mathbf{f}_{i}=\frac{\Omega_\mathrm{K}}{\mathrm{St}_{{i}}} (\mathbf{u}-\mathbf{v}_{{i}})\,.
\end{equation}
Here, the Stokes number of a specific dust species, $\mathrm{St}_{{i}}$, is a dimensionless parameter, incorporating the dynamical relevant quantities of a spherical dust grain at a certain location in the disk:
\begin{equation}
	\mathrm{St}_{{i}} = \frac{\pi}{2}\frac{a_{{i}} \rho_\mathrm{mat}}{\Sigma_\mathrm{g}}\, , 
\end{equation} 
with $a_{{i}}$ the radius of the dust grain and $\rho_\mathrm{mat}$ its intrinsic material density. 

The vector $\mathbf{j}_{i}$ appearing in Equation~(\ref{eq:contdust}) symbolizes the diffusion flux of a dust fluid due to a gradient in its concentration and is expressed by (following \citet{1984ApJ...287..371M}):
\begin{equation}
	\mathbf{j}_{{i}} = -D_{{i}}(\Sigma_\mathrm{g}+\Sigma_{{i}})\, \mathbf{\nabla} \left( \frac{\Sigma_{i}}{\Sigma_\mathrm{g}+\Sigma_{i}}\right)\,.
	\label{eq:diffflux}
\end{equation}
The diffusion in our model is employed to mimic stochastic kicks exerted onto the dust grains by the surrounding turbulent gas fluid. This implies that the diffusion coefficient of each dust species is a function of the turbulence, which in turn is expressed by the kinematic viscosity, $\nu$. We apply the relationship \citet{2007Icar..192..588Y} found for radial diffusion:
\begin{equation}
	D_{i} = \nu\frac{1+4\mathrm{St}_{\rm i}^2}{\left(1+\mathrm{St}_{\rm i}^2\right)^2}\,,
\end{equation}
in both radial and azimuthal direction.
A detailed description and numerical tests of the diffusion implementation are presented in Appendix~\ref{appendix}.

\subsection{Fiducial Model}
Here, the parameters used in the fiducial model are introduced and shortly justified.
\begin{table}
	\centering
	\caption{\textrm{Set of parameters for fiducial hydrodynamical simulations.}}
	\label{tab:para}
	\begin{tabular}{l l r }
		\hline\hline
		Stellar mass \qquad  \qquad &$m_\ast\,[\mathrm{M}_\odot]$  \qquad & $1$ \\
		Planet mass \qquad  \qquad &$m_\mathrm{p}\,[\mathrm{M}_\oplus]$  \qquad & $17$ \\
		Planet semi-major axis \qquad  \qquad &$r_\mathrm{p}\,[\mathrm{AU}]$  \qquad & $30$ \\
		Planet smoothing parameter \qquad  \qquad \quad& $b$ \quad \qquad & $0.6$ \\[4pt]
		Surface density at $r_\mathrm{p}$ \qquad  \qquad &$\Sigma_{\rm g}\,[\mathrm{g}\mathrm{cm}^{-2}]$  \qquad & $5$ \\
		Surface density slope \qquad  \qquad&$\sigma$   \qquad& -0.5 \\
		Temperature slope \qquad  \qquad&$\xi$   \qquad& -1.0 \\
		Aspect ratio at $r_\mathrm{p}$ \qquad  \qquad&$h$   \qquad& 0.05 \\
		Viscosity parameter \qquad  \qquad & $\alpha$   \qquad& $10^{-5}$ \\[4pt]
		Number of dust species \qquad  \qquad \quad& $\mathrm{N}_\mathrm{dust}$ \quad \qquad & $5$ \\
		Minimum particle size \qquad  \qquad \quad& $a_\mathrm{min}\,[\mathrm{mm}]$ \quad \qquad & $10^{-2}$ \\
		Maximum particle size \qquad  \qquad \quad& $a_\mathrm{max}\,[\mathrm{mm}]$ \quad \qquad & $1$ \\
		Material density \qquad  \qquad \quad& $\rho_\mathrm{mat}\,[ \mathrm{g} \,\mathrm{cm}^{-3}]$ \quad \qquad & $2$ \\
		Total dust-to-gas ratio \qquad  \qquad \quad& $\varepsilon$ \quad \qquad & $0.01$ \\[4pt]
		Number of radial cells \qquad  \qquad \quad& $\mathrm{N}_r$ \quad \qquad & $600$ \\
		Number of azimuthal cells \qquad  \qquad \quad& $\mathrm{N}_\varphi$ \quad \qquad & $1200$ \\
		\hline
	\end{tabular}
\end{table}
All relevant parameters are listed in Table~\ref{tab:para}. We choose a star of solar mass and a Neptune mass planet located at $r_\mathrm{p}=30\,\mathrm{AU}$ and adopt typical values at this location for the aspect ratio and surface density. An important parameter for hydrodynamical processes is the level of viscosity, typically quantified by the dimensionless stress, $\alpha$. Following the idea that angular momentum transport is rather caused by disk winds launched in upper regions of the disk than through viscous processes, we investigate the effects of the planet in a low-viscosity disk with $\alpha=10^{-5}$.\\
For all our models, we use a dust-to-gas ratio of $\varepsilon=0.01$ and assume that the number density of dust grains follows a power-law $n(a)\propto a^{-\gamma}$. We choose a conservative value for  the power-law index, $\gamma=3.5$, and a maximum grain size, $a_\mathrm{max}=1\,\mathrm{mm}$, that implicitly assumes that grain growth is an ongoing process in outer regions of the disk but limited due to particle drift, bouncing and fragmentation \citep[see e.g.][for a review of laboratory results]{2018SSRv..214...52B}. This is motivated by recent interpretations of PPD observations \citep[see e.g.][]{2019MNRAS.486L..63R,2019ApJ...877L..18Z} and, assuming the grains to be compact spheres, also represents roughly the largest observable grain size for ALMA observations. The minimum particle size of $a_\mathrm{min}=10\,\mu\mathrm{m}$ is unrealistically large, however, as long as $a_\mathrm{max} \gg a_\mathrm{min}$ this parameter does not influence the dynamics \citep{2007ApJ...671.2091G} and smaller particles are taken to perfectly follow the profile of the gas due to their strong coupling. We stress, that the grain sizes are kept fixed in our simulations and are therefore not accounting for any size evolution.  

\subsection{Numerical Setup}
The hydrodynamical simulations in this work make use of the publicly available code FARGO3D\footnote{\href{http://fargo.in2p3.fr/}{http://fargo.in2p3.fr/}} \citep{2016ApJS..223...11B} with multiple fluids \citep{2019ApJS..241...25B}, enabling the study of both the gas and dust structures by including several dust species as pressureless fluids and solving Equations~(\ref{eq:contgas})-(\ref{eq:NS-dust}) numerically.
In all our runs presented in this work we use the FARGO algorithm \citep{2000A&AS..141..165M}.
The grid is spread linearly in azimuth and logarithmically in radius, using a number of cells, $\mathrm{N}_\varphi \times \mathrm{N}_r = 1200 \times 600$, spread over the full azimuthal angle and $12\,\mathrm{AU}< r < 75\,\mathrm{AU}$.   
The disk temperature, $T$, and the gas and dust densities, $\Sigma$, are assumed to initially follow a power-law structure:
\begin{equation}
	T\propto r^{\xi}\,, \qquad \quad \Sigma \propto r^{\sigma}\,,
\end{equation}
with $\xi=-1.0$ and $\sigma=-0.5$. The dust is modelled by five distinct fluids interacting with the gas (not explicitly interacting with each other) and only differ in the grain size that the fluid represents. The different grain sizes are spaced logarithmically between the minimum grain size and the maximum grain size. This -- together with the size dependency of the particle number density $n(a)\propto a^{-3.5}$ -- translates into a depency of each dust component surface density of $\Sigma_{i}(a) \propto \sqrt{a}$. The initial density structure of each dust species is then fully determined by demanding the total dust-to-gas ratio to be $\varepsilon = 0.01$ everywhere in the disk.

We set the boundary conditions to steady state drift solutions that are given in \citet{2019ApJS..241...25B} for gas and an arbitrary number of dust species. These solutions account for the momentum exchange between gas and an arbitrary number of dust species. The drift velocity of a dust species is therefore dependent on the motion of the other dust species, since the dust's feedback changes the gas background flow. To prevent reflections of the planet's wake at the boundaries, we damp density and velocity fields close to the boundaries according to \citet{2006MNRAS.370..529D} following the prescription in \citet{2016ApJ...826...13B}.

\section{Hydrodynamical Results}\label{sec:hydro}
\subsection{Planet on fixed orbit}\label{subsec:fixed}
\begin{figure*}
	\centering
	\includegraphics[width=\textwidth]{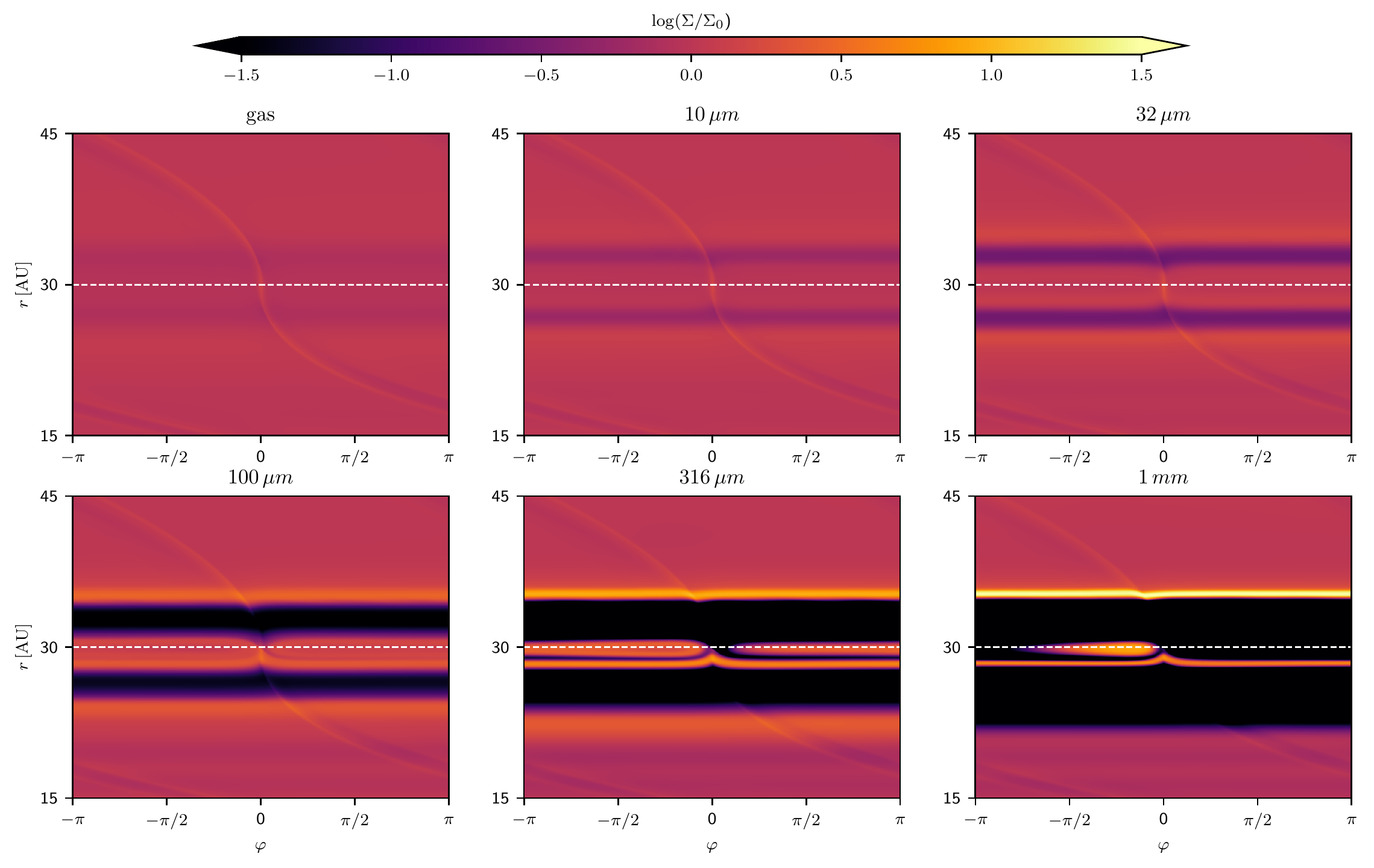}
	\caption{Normalized surface density of different dust species in the case of a planet on a fixed (non-migrating) orbit. The planet is located at $r=30\,\mathrm{AU}$ and the density is shown after $t=300\,\mathrm{orbits}$. All relevant parameters are listed in Table~\ref{tab:para}.}
	\label{fig:nomig2d}
\end{figure*}

In Figure~\ref{fig:nomig2d} we present the density structure of gas and several dust species, perturbed by a planet of Neptune mass in the case of $\alpha=10^{-5}$. This fiducial model uses the parameters that are summarized in Table~\ref{tab:para}.  

Prominently, one can observe two gaps in the dust structures. One possible explanation for this phenomenon can be found in \citet{2001ApJ...552..793G}, who calculated that density waves that are launched by the gravitational pull of an embedded planet eventually turn into shocks. Hence, they cause a local perturbation of the gas pressure profile, that affects the radial dust velocities and can cause gaps. As a result dust can accumulate at the gap edges as shown by \citet{2017ApJ...843..127D} in hydrodynamical simulations including gas and one dust size.
The authors concluded from this, that a single planet can cause an observable triple-ring structure: One ring outside the outer gap, one ring being trapped in the planet's horseshoe orbit and one ring inside to the inner gap.

The results presented in Figure~\ref{fig:nomig2d} highlight, that the obtained structure depends on the regarded grain size. While the ring created outside the planet's orbit is located at the same radial distance to the star for all five dust species, the accumulation of solids in the horse-shoe orbit and radially inwards of the planet's location show distinct features for different dust sizes. 

\begin{figure}
	\centering
	\includegraphics[width=\columnwidth]{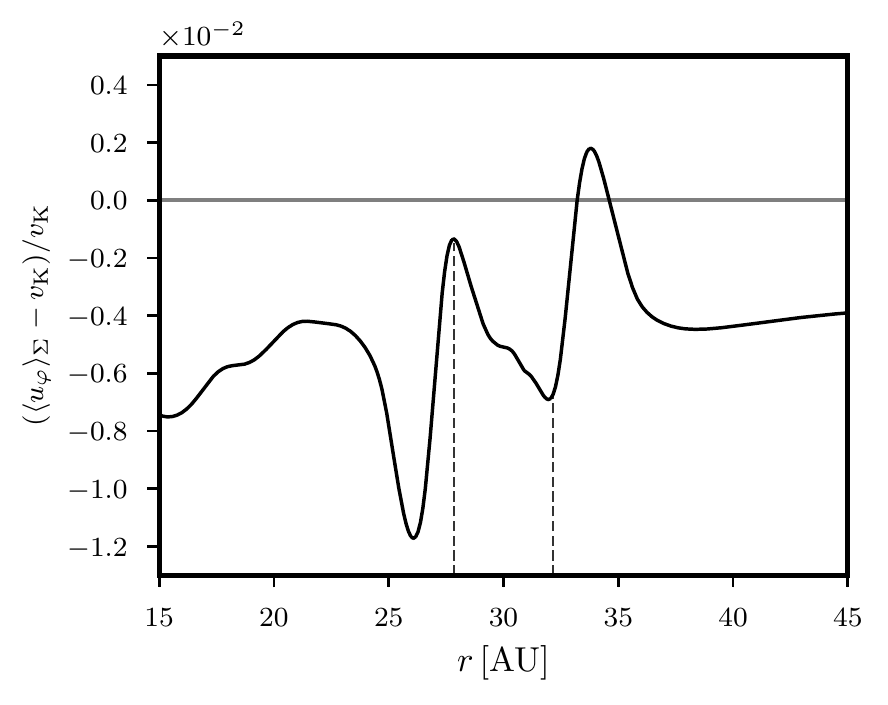}
	\caption{Azimuthal velocity of the gas in its density-weighted azimuthal average, $\langle u_\varphi \rangle_\Sigma = \langle u_\varphi \Sigma_\mathrm{g}\rangle_\varphi / \langle\Sigma_\mathrm{g} \rangle_\varphi$. The azimuthal velocity is plotted in its relative deviation from the Keplerian profile.  It corresponds to the same setup as Figure~\ref{fig:nomig2d}. The vertical dashed lines mark the positions calculated by \citet{2001ApJ...552..793G}, where shocks are predicted to develop.}
	\label{fig:azvel}
\end{figure}

The distribution of dust can be best understood by recalling that small grains are tightly coupled to the dynamics of the gas, which means that their radial flow follows the radial flow of the gas. The radial velocities of decoupled dust grains, however, are most strongly affected by the head- or tailwind they experience from the gas, i.e. by the deviation of the azimuthal gas velocity from the Keplerian speed. Figure~\ref{fig:azvel} shows this deviation in its azimuthal average, weighted by the gas surface density and normalized to the Keplerian speed. By analyzing these velocity profiles one can understand the created dust distribution in Figure~\ref{fig:nomig2d}: The gravitational pull of the planet causes the azimuthal velocity of the gas to become super-Keplerian in the outer disk. This imposes a tailwind onto decoupled dust grains, hence a positive torque causing them to drift outwards. The outer ring is formed at the location, where the azimuthal velocity of the gas is Keplerian and the accumulation is the most pronounced for the most decoupled species. Looking more closely at Figure~\ref{fig:azvel} it becomes apparent that there are two such locations in the outer disk, narrowly spaced. The inner location of Keplerian velocity is, however, an unstable equilibrium. Partly decoupled particles that are located a bit further outside are transported outwards and if they are located slightly further inside, their drift is inwardly directed. The outer radius of Keplerian rotation is in contrast a point of accumulation.

As mentioned before, small dust is following the radial velocity of the gas. Consequently, the small dust species are not completely halted at the pressure maximum and no ring is formed. 
In the inner system the azimuthal velocity of the gas shows a local maximum but it does not turn super-Keplerian anywhere. As a consequence, the radial drift of the dust is merely slowed down but not reversed and the largest dust grains (that are not being replenished due to trapping in the outer disk) begin to be evacuated from this area. This is not the case for dust species that are coupled more strongly. For them the local maximum is enough to create a congestion of dust at these radii that appears as a ring in its density structure. Due to the different dynamical behavior of distinct dust sizes, this ring is located at slightly shifted radii for different species. 

\subsection{The effect of migration}
The uncertainty in the migration behavior of an embedded planet motivates us to keep this quantity as an input parameter of the simulations. This means the planetary motion is entirely prescribed, allowing us to study how the PPD adapts to different migration scenarios. The prescription is set by updating the planet position at every time step:
\begin{eqnarray}
r_\mathrm{p}^{n+1}\,=\,r_\mathrm{p}^n + \dot{r}_\mathrm{p}^{n}\mathrm{d}t \,&,& \label{eq:update}\\ [4pt]	
\dot{r}_\mathrm{p}^{n}\, =\, \frac{r_\mathrm{end} -r^n}{t_\mathrm{end}-t^n} \, &,& \label{eq:migrate}
\end{eqnarray} 
where $r_\mathrm{end}$ is the distance to the star at which the planet ends up after a time of $t_\mathrm{end}$. In all of our simulations we model migration for $t_\mathrm{end}=300\,\mathrm{orbits}$ and set $r_\mathrm{end} = 30\,\mathrm{AU}$, such that at the end of the simulations different scenarios are directly comparable to each other. The migration rates are therefore completely controlled by setting the orbital distance at which the planet is initialized, $r_\mathrm{p,0}=r_\mathrm{p}(t=0)$:
\begin{equation}
	\dot{r}_\mathrm{p} = \frac{r_\mathrm{end}-r_\mathrm{p,0}}{t_\mathrm{end}}\,.
\end{equation}   

As a point of reference, we take the so-called Type I migration rate by \citet{2002ApJ...565.1257T} in its two-dimensional representation, which is one of the standard predictions for a planet not massive enough to open a gap in the gas structure. For parameters used in the fiducial model this results in a migration rate of $\dot{r}_\mathrm{p}=-6.1\times10^{-5}\mathrm{AU}/ \mathrm{yr}$ which is equivalent to $\dot{r}_\mathrm{p}=-3.4\times10^{-4}r_\mathrm{p}/ \mathrm{orbit}$. We use the absolute value of this speed as a reference migration speed in negative and positive direction and investigate one slower and one faster mode, respectively. Different migration scenarios are summarized in Table~\ref{tab:paramig}.

\begin{figure}
	\centering
	\includegraphics[width=\columnwidth]{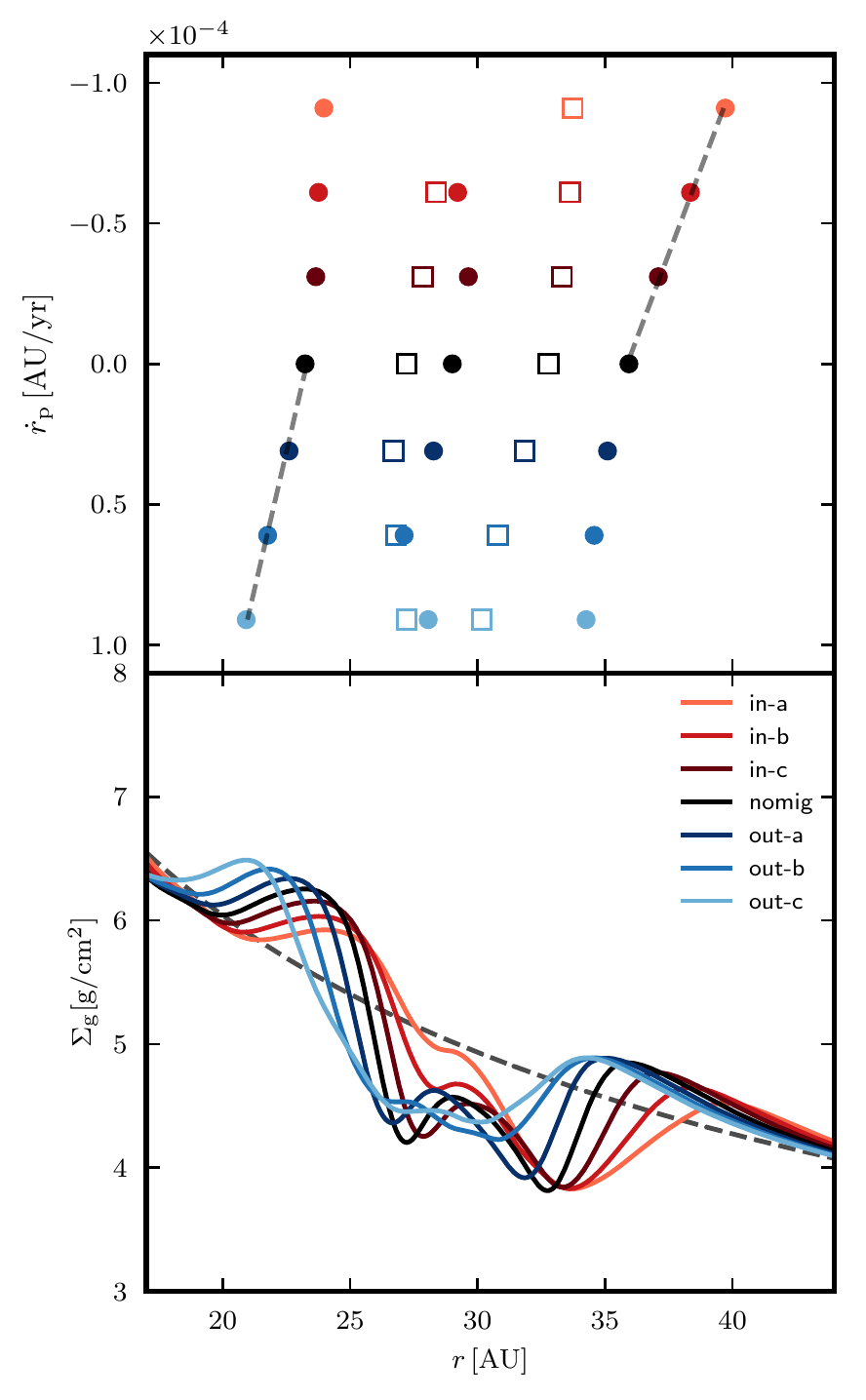}
	\caption{Radial structure of gas surface density for different migration speeds of the planet. The lower panel shows the radial density profiles in the vicinity of the planet, which is located at $r_\mathrm{p}=30\,\mathrm{AU}$ in all the cases. The dashed line shows the initial gas distribution. The upper panel shows the locations of maxima (circles) and minima (squares) of the averaged gas surface density as a function of migration speed.  Their colors correspond to the equivalent color in the legend of the lower panel. The dashed lines show linear fits to the locations of the outer maximum (Equation~\ref{eq:r_out}) and inner maximum (Equation~\ref{eq:r_in}) for inward migration and outward migration, respectively. Note, that in the case of fast inwards migration the central gas density maximum vanishes.}
	\label{fig:fid-gas}
\end{figure}

\begin{table}
	\centering
	\caption{\textrm{Set of explored migration rates.}}
	\label{tab:paramig}
	\begin{tabular}{l c c }
		\hline\hline
		Name \qquad \quad \quad & $\dot{r}\,[10^{-5}\times\mathrm{AU}/\mathrm{yr}]$ \quad \quad \qquad & $R_\mathrm{0} [\mathrm{AU}]$\\
		\hline
		in-a \qquad  \qquad &$-9.1$ &$34.5$   \\
		in-b \qquad  \qquad &$-6.1$ & $33$  \\
		in-c  \qquad  \qquad &$-3.1$ & $31.5$ \\
		fixed \qquad  \qquad &0.0 & $30$  \\
		out-a \qquad  \qquad &$3.1$ & $28.5$   \\
		out-b \qquad  \qquad &$6.1$ & $27$   \\
		out-c \qquad  \qquad & $9.1$  &$25.5$  \\
		\hline
	\end{tabular}
\end{table}

Figure~\ref{fig:fid-gas} shows the radial structure of the gas surface densities for different cases of planet migration. In all the snapshots the planet is located at $r = 30\,\mathrm{AU}$ after an evolution of $t=300\,\mathrm{orbits}$. 
The perturbations of the gas density are only small compared to the initial profile. We note that the planet compresses the density maximum it is migrating towards, while the one in the opposite direction is diminished and broadened, and its location is shifted to radii more distant to the planet location. This is in accordance with the assessment done in \citet{2019MNRAS.482.3678M} and \citet{2019AJ....158...15P}. For largest absolute migration rates, the maximum in the horseshoe region starts to dissolve and the gap depletion is less pronounced. 

It is clearly visible, that in the case of an inward migrating planet the outer maximum is located the further outwards, the faster the planet is migrating. Equivalently, for an outwards migrating planet the inner maximum is shifted the further inwards, the faster the planet is migrating. 

In the right panel of Figure~\ref{fig:fid-gas} we fitted the distance of the relevant maxima to the planet's orbit with linear functions shown as dashed grey lines. They are given by:
\begin{eqnarray}\label{eq:r_out}
	r_\mathrm{out}(\dot{r}_\mathrm{p}) = 35.9\,\mathrm{AU}  - 4.16\times10^{4}\mathrm{yr} \times\dot{r}_\mathrm{p}\left[\frac{\mathrm{AU}}{\mathrm{yr}}\right]\,&,&\\[4pt]
		\label{eq:r_in}
		r_\mathrm{in}(\dot{r}_\mathrm{p}) = 23.3\,\mathrm{AU} - 2.56\times10^{4}\mathrm{yr} \times\dot{r}_\mathrm{p}\left[\frac{\mathrm{AU}}{\mathrm{yr}}\right]\,&.&
\end{eqnarray} 

\subsection{Optical depth}\label{subsec:optdepth}
In an attempt to make a simple estimate of an average dust profile that is meaningful for observations, we calculate the optical depth of the simulated disk for a specific observational wavelength, $\lambda_\mathrm{obs}=1.3\,\mathrm{mm}$:
\begin{equation}
\tau_\lambda = \sum_{i=1}^{\mathrm{N}_\mathrm{dust}} \kappa_{{i}}(\lambda_\mathrm{obs})\Sigma_{i}\,,
\end{equation} 
where $\kappa_{{i}}(\lambda_\mathrm{obs})$ are rough estimates \citep{1997MNRAS.291..121I} for the absorption opacity of dust species $i$:
\begin{equation}
\quad \kappa_{{i}}(\lambda_\mathrm{obs}) = \kappa^\mathrm{geo}_{i}\times \min\left(1,\frac{2\pi a_{i}}{\lambda_\mathrm{obs}}\right)\,,\quad \kappa^\mathrm{geo}_{i} = \frac{3}{4a_{i}\rho_\mathrm{mat}}\,.
\end{equation}

\begin{figure}
	\centering
	\includegraphics[width=\columnwidth]{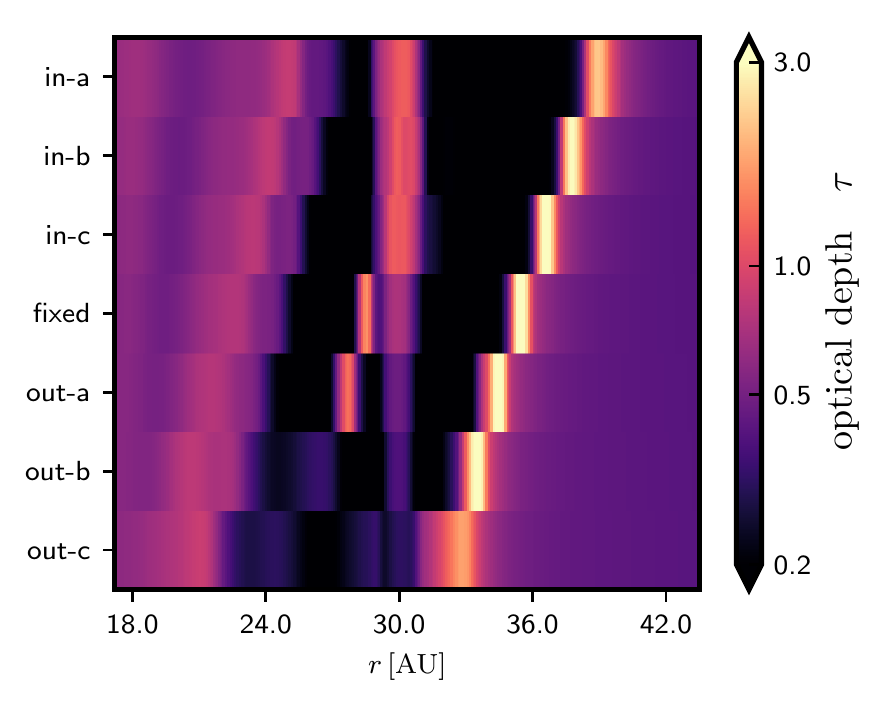}
	\caption{The color code traces the optical depth for a wavelength of $\lambda_\mathrm{obs}=1.3\,\mathrm{mm}$ in azimuthal average for the fiucial model. Different migration rates are spread out on the y-axis, while the x-axis shows the radial direction around the planet's position after 300 orbits, $r_\mathrm{p}=30\,\mathrm{AU}$.}
	\label{fig:fid-kdens}
\end{figure}

Figure~\ref{fig:fid-kdens} shows the calculated optical depth for different scenarios of planet migration. Similar to the effect observed in the gas structure, the locations of dust accumulation become more and more asymmetric towards each other when increasing the absolute value of the migration rate.

\section{Synthetic Observations}\label{sec:synt-observ}
\subsection{Radiative Transfer and Image Synthesis}
Synthetic image predictions based on the hydrodynamic simulations are obtained via a radiative transfer calculation. Here we use the publicly available code RADMC3D\footnote{\href{http://www.ita.uni-heidelberg.de/~dullemond/software/radmc-3d/index.html}{http://www.ita.uni-heidelberg.de/$\sim$dullemond/software/radmc-3d/index.html}} \citep[version 0.41,][]{2012ascl.soft02015D}.

For running the radiative transfer calculations, we need to transform the output of the two-dimensional hydrodynamical simulations into a vertically expanded, three-dimensional structure. We therefore expand the grid into spherical coordinates, opening an angle of $\Delta\theta=\pi/16$ upwards from the midplane in 32 polar grid cells that are logarithmically refined towards the midplane. This extent contains roughly four vertical gas scale heights of the disk. We assume a Gaussian shape for the vertical distribution of the densities and demand that the vertically integrated mass of each species is conserved\footnote{Note, that also the vertical boundary $z_\mathrm{max}$ is changing with $r$: $z_\mathrm{max}=\mathrm{sin}(\Theta_\mathrm{min})r$.}. Thus, we enforce:
\begin{equation}
	\int_{-z_\mathrm{max}}^{z_\mathrm{max}}\rho_{0,{i}}(r) \exp\left(-\frac{z^2}{2H_{i}^2}\right)\, \mathrm{d}z = \Sigma_i \,,
\end{equation}
which is fulfilled, if
\begin{equation}\label{eq:rho0}
	\rho_{0,{i}}(r) = \frac{\Sigma_i(r)}{\sqrt{2\pi}H_{i}(r)}\times\mathrm{erf}^{-1}\left(\frac{\mathrm{z}_\mathrm{max}}{\sqrt{2}H_{i}(r)}\right)\,,
\end{equation}
where $\mathrm{erf}()$ is the error function.
The vertical structure of the disk is therefore defined by the pressure scale height of the disk. We emphasize that Equation~\ref{eq:rho0} deviates from the standard analytical expression for $\rho_0$ by the contribution of the error function, which accounts for the finite vertical grid.
For the dust grains we consider vertical spreading through diffusion with diffusion coefficient, $D_z=\nu/\mathrm{Sc}_z$, that assumes the diffusion to be linked to the viscosity of the gas through the vertical Schmidt-number, $\mathrm{Sc}_z$. Following the standard diffusion model \citep{1995Icar..114..237D}, the vertical scale height of the dust is defined as:
\begin{equation}\label{equ:Hd2}
H_{i} =  \sqrt{\frac{\tilde{\alpha}}{\tilde{\alpha}+\mathrm{St}_{i}}} H_\mathrm{g}\, ,
\end{equation}
where we define $\tilde{\alpha}:= \alpha/\mathrm{Sc}_z$. The Stokes number, $\mathrm{St}$, is strictly positive, which limits the dust scale height to be smaller than the gas scale height.
In all our models we assume a factor of $\mathrm{Sc}_z = 0.1$ for the vertical spreading of the dust fluids.
To sample variations in the opacity curve of the dust mix, we decompose the size range of the five simulated dust species (symbolized by $i$) into forty sub-bins each (symbolized by $j$). These sub-bins are evenly spaced in logarithmic space, such that for each $i$:
\begin{equation}
	\Sigma_j\propto a_{j}^{0.5}\,, \quad \mathrm{and}, \quad \sum_{j=1}^{40} \Sigma_{j} = \Sigma_i\,.
\end{equation}

Further, in the outputs from hydrodynamical simulations, all the fields describing the gas and dust fluids are limited to the radial domain of the simulation $[0.4r_\mathrm{p},2.5r_\mathrm{p}]$. Especially the inner boundary can lead to artificial ring observations, because the sudden edge of density at this location that is exposed to direct stellar illumination heats up to high temperatures. To reduce the effect this inner boundary has on the observed fluxes in the region of interest, i.e. around the planet's orbit, we take two precautions: first, we interpolate the dust density structure for the disk to extent to radii further inside to increase the distance between the gap structure and the disk edge. Secondly, we introduce a small unresolved inner disk around the star that blocks parts of the direct illumination. The interpolation is done by using the unperturbed density power-law defined as the initial condition of the hydrodynamical simulations and extend it to an inner value of $r=2\,\mathrm{AU}$. Additionally, we smooth the inner edge by applying an exponential cut-off.

The inner dust disk is introduced between $0.05\, \mathrm{AU}$ and $0.15 \, \mathrm{AU}$ before running the radiative transfer calculations. Here, seven different dust species logarithmically spaced between $a_\mathrm{min}=1\,\mu\mathrm{m}$ and $a_\mathrm{max}=1\,\mathrm{mm}$ are distributed between $0.05\,\mathrm{AU}$ and $0.15\,\mathrm{AU}$ with: 
\begin{equation}
	\rho_{i} = \frac{\Sigma_{i}}{\sqrt{2\pi}H_{i}}\exp\left(-\frac{z^2}{2H_{i}^2}\right)\,,
\end{equation}
where the surface density of each individual dust species is set to
\begin{equation}
	\Sigma_{i}= 0.2 \frac{\mathrm{g}}{\mathrm{cm}^2}\times\left(\frac{r}{0.1\mathrm{AU}}\right)^{-0.5}\,,
\end{equation}
and the aspect ratio is the same as in the outer disk, but reduced by a factor of 0.6.

The central star is assumed to be of Solar type, with a mass of $M_\ast=1\,M_\odot$, a radius of $R_\ast=1\,R_\odot$ and a surface temperature $T_\ast=6000\,\mathrm{K}$. 
The disk temperature is calculated by the Montecarlo-process of RADMC3D with $n_\mathrm{phot}=10^8$ photon packages emitted by the star. To account for different grain properties we assume all dust species to be a mixture of $30\%$ amorphous carbon and $70\%$ silicates. The optical constants of the mix are computed by applying the Bruggeman rules. The material density of the dust in the mix is $1.7\,\mathrm{g}\,\mathrm{cm}^{-3}$. Then, the Mie theory was used to calculate the mix's opacities. The midplane temperature achieved with this procedure is typically of the order of $25\,\mathrm{K}$ at the location of the planet.

To infer whether or not the produced features in the PPD are observable by a state-of-the-art telescope, we simulate an ALMA observation using the simobserve function of the casa-5.5.0 software package \citep{2007ASPC..376..127M}. To obtain high spatial resolution we choose a long baseline configuration ("alma.cycle5.9") and integrate the flux for ten hours.  Additionally, we integrate for two hours in a short baseline configuration ("alma.cycle5.7") to complete the UV-plane with short spacings. The angular resolution achieved with this is $0\farcs024\times 0\farcs032$. We assume the observed object to be at a distance of $d=100\,\mathrm{pc}$ and to be observed face-on, for which angular distances translate according to $0\farcs01 \,\hat{=}\, 1\,\mathrm{AU}$.
\begin{figure*}
	\centering
	\includegraphics[width=\textwidth]{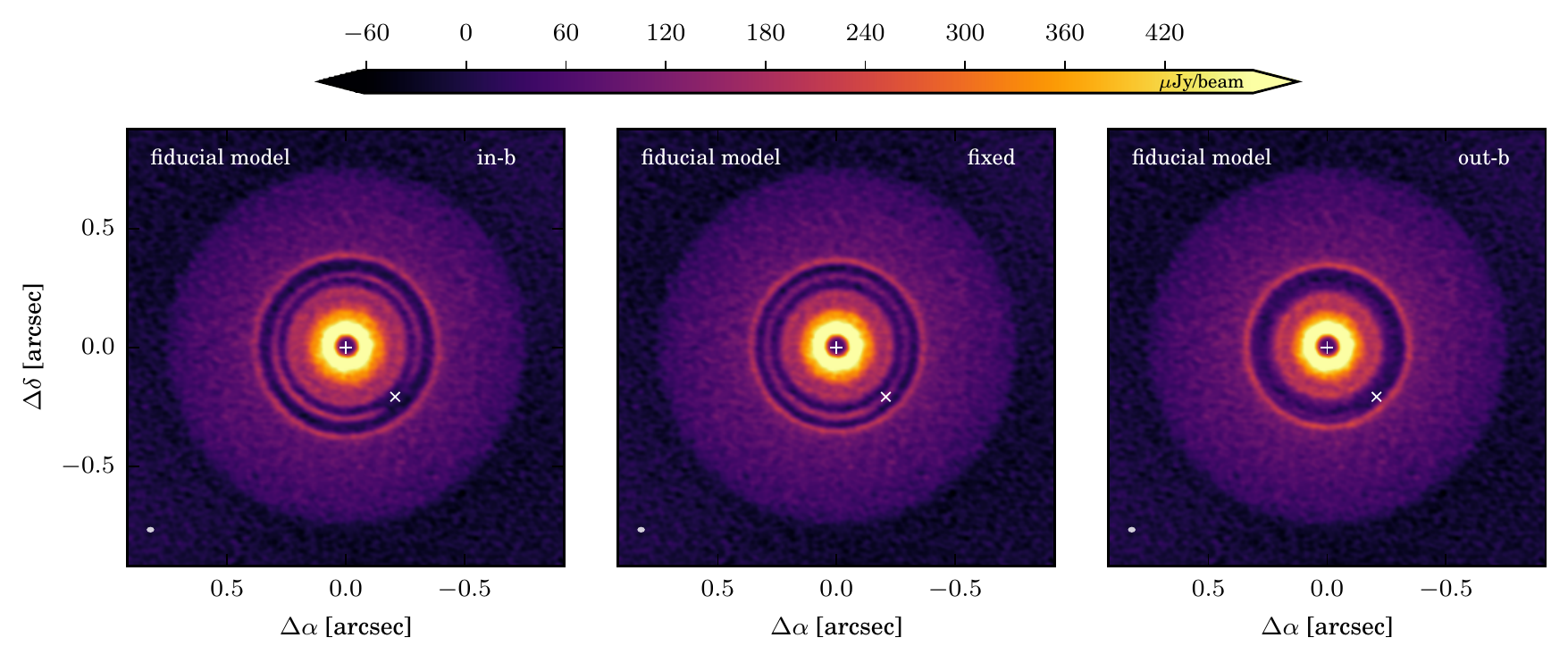}
	\caption{Synthetic images using CASA for an observing wavelength of $\lambda_\mathrm{obs}=1.3\,\mathrm{mm}$. The left panel shows an inwards migrating planet ("in-b"), the central panel a planet on a fixed orbit ("fixed") and the right panel a planet migrating outwards ("out-b"). The obtained beam size from the simulated ALMA-configuration is $0\farcs024\times0\farcs032$, which is visualized as a white ellipse in the bottom left corner. The white plus sign shows the position of the central star, for which physical properties are set to Solar values for temperature and size and is placed at a distance of $100\,\mathrm{pc}$ to the observer. The white cross shows the position of the planet, which is the same for all three examples, $r_\mathrm{p}=30\,\mathrm{AU}\,\hat{=}\, 0\farcs3$. }
	\label{fig:synthetic}
\end{figure*}

Figure~\ref{fig:synthetic} shows the predicted images for three scenarios of the fiducial model. In the left picture the planet is migrating inwards ("in-b"), in the center the planet is not migrating at all ("fixed") and on the right the planet is migrating outwards ("out-b"). The structures are different and distinguishable  for the three scenarios. While in the first two cases the triple-ring structure is prominent, in the case of fast outwards migration the image appears more blurred. This is confirmed by the structure of the estimated optical depth for this case shown in Figure~\ref{fig:fid-kdens}. The contrast between the central ring and the gaps is small in this case. For the case of inwards migration, the asymmetric spacing between the rings can be easily noticed, something that is absent in the case of a fixed planet.

\subsection{Azimuthally-averaged intensity profile}
As a last step we want to investigate, whether the ring locations found in the hydrodynamical simulations in Section~\ref{sec:hydro} can in fact be extracted from the synthetic observations. For this, we average the intensity of the synthetic observations azimuthally and plot it as a function of radius.
\begin{figure}
	\begin{minipage}[c]{\columnwidth}
		\centering
		\includegraphics[width=\columnwidth]{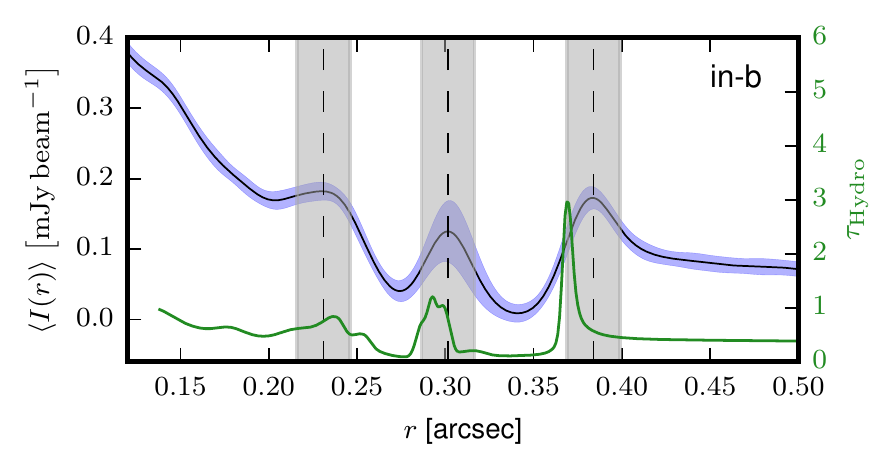}
	\end{minipage}
	\\[-17pt]
	\begin{minipage}[c]{\columnwidth}
		\centering
		\includegraphics[width=\columnwidth]{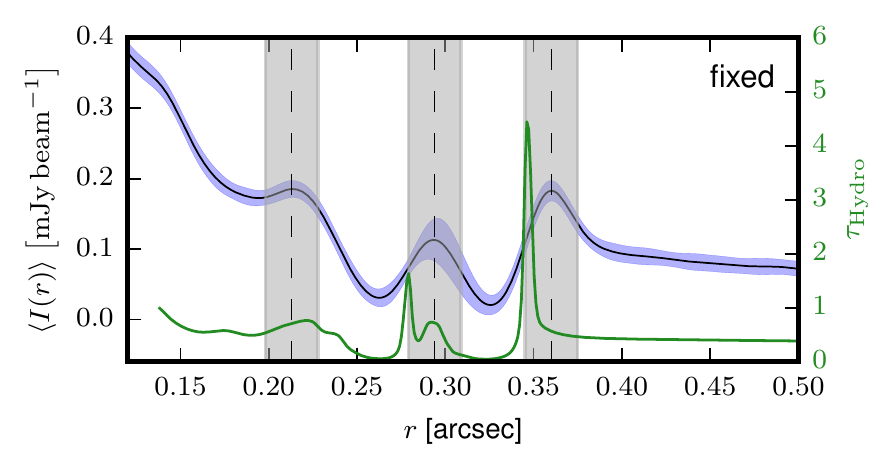}
	\end{minipage}
	\\[-17pt]
	\begin{minipage}[c]{\columnwidth}
		\centering
		\includegraphics[width=\columnwidth]{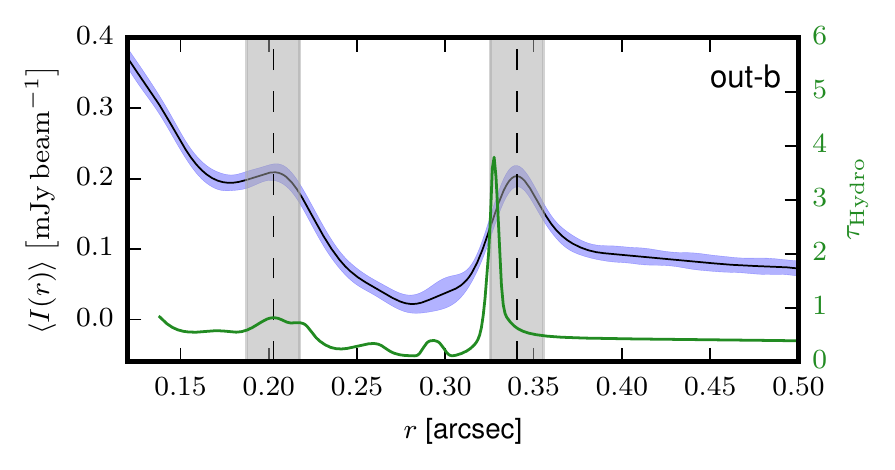}
	\end{minipage}
	\caption{The black line shows the azimuthally-averaged surface brightness profiles of different migration cases. The blue-shaded area displays the standard deviation of the azimuthal average. The forestgreen line shows the direct comparison to the optical depths estimated in Section~\ref{subsec:optdepth}.}
	\label{fig:azav-int}
\end{figure}
Figure~\ref{fig:azav-int} shows the radial profiles of the synthetic images presented in Figure~\ref{fig:synthetic}. This also allows the comparison to the approximated optical depth which is added to the figures in forestgreen. It becomes obvious, that while the general asymmetry of the ring structure persists after the radiative transfer, the exact location of averaged intensity and optical depth is slightly shifted. This is especially true for the position of the inner ring. One reason for this shift to smaller radii in the synthetic images is the inwardly increasing temperature. 

\section{Influence of key parameters}\label{sec:parameters}
In this section we want to discuss how relative ring spacings depend on important parameters of the PPD or the planet.
\paragraph{Viscosity}
A low level of viscosity is a requirement to form the triple-ring structure that we intend to investigate. The fiducial value of $\alpha=10^{-5}$ implies a viscosity that does not dominate most of the dynamical processes within the disk and indeed we hardly find any significant modification of the structures neither for inviscid simulations nor for $\alpha=10^{-4}$. In fact, the ring and gap locations stay almost the same and there are only slight changes in the density contrast between maxima and minima. If one increases the viscosity further, however, the triple-ring structure is lost.
\begin{figure*}
	\centering
	\includegraphics[width=\textwidth]{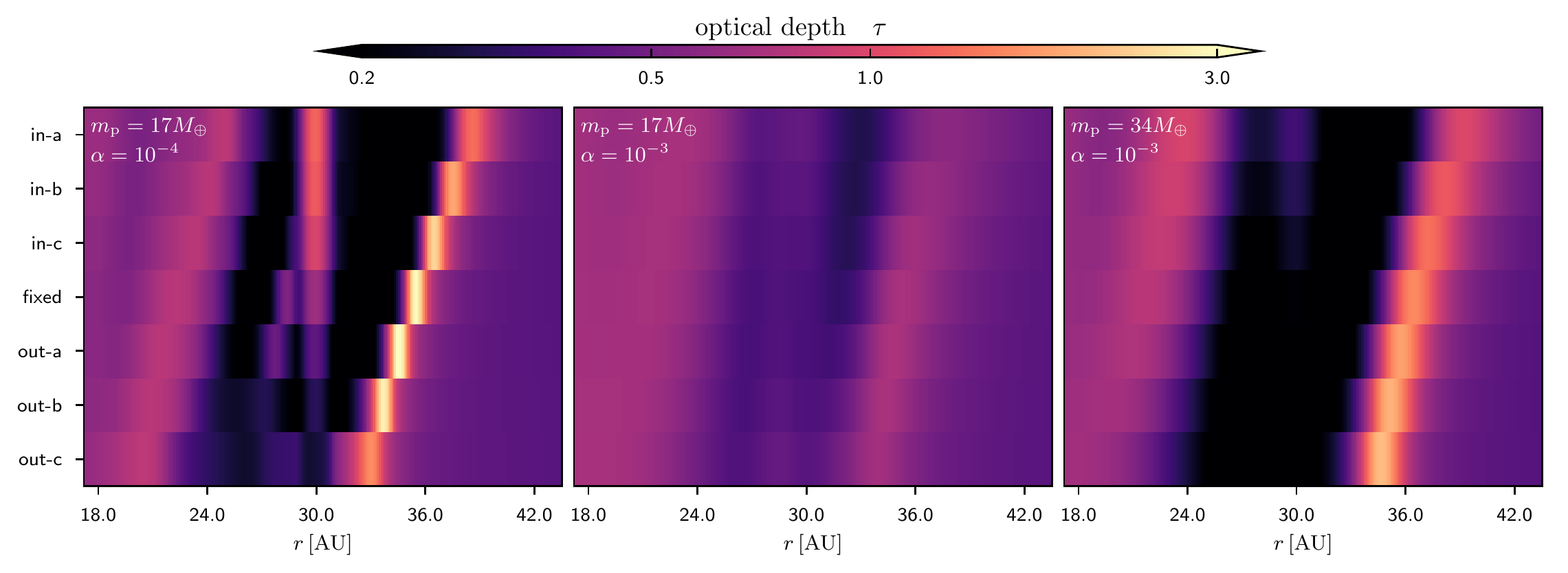}
	\caption{Azimuthal average of optical depth after 300 orbits for $m_\mathrm{p}=17M_\oplus$ and $\alpha=10^{-4}$ (left panel), $m_\mathrm{p}=17M_\oplus$ and $\alpha=10^{-3}$ (central panel) and $m_\mathrm{p}=34M_\oplus$ and $\alpha=10^{-3}$ (right panel). Except for planet mass, $m_\mathrm{p}$, and level of viscosity, $\alpha$, all model parameters are those of the fiducial setup described in Table~\ref{tab:para}.}
	\label{fig:visc}
\end{figure*}
This can be seen in Figure~\ref{fig:visc}. The left panel shows that an increase of the viscosity to $\alpha=10^{-4}$ qualitatively produces the same results as in the fiducial model displayed in Figure~\ref{fig:fid-kdens}, the central panel of Figure~\ref{fig:visc} shows that when increasing the viscosity further, however, a planet of $m_\mathrm{p}=17M_\oplus$ is not strong enough to cast a perceivable structure. When increasing the planet mass to $m_\mathrm{p}=34M_\oplus$ and keeping the viscosity coefficient at $\alpha=10^{-3}$, the right panel shows that the gap and ring structure is once more created but lacks the characteristic central accumulation in the horseshoe region. We find that an inwards migrating planet enhances the optical depth in the inner ring and an outwards migrating planet in the outer one. The effect of migration on gas and dust in this case of a higher level of viscosity $\alpha > 10^{-4}$ has been studied by \citet{2019MNRAS.482.3678M} with application to synthetic observations in \citet{2019MNRAS.485.5914N}.The optical depth structure in the right panel of Figure~\ref{fig:visc} suggests qualitative agreement between our evalutaion of the high viscosity case and the results of \citet{2019MNRAS.485.5914N} who showed that for faster inward migration the outer ring tends to disappear while the inner accumulation grows more pronounced.

\paragraph{Aspect ratio and planet mass}
\begin{figure*}
	\centering
	\includegraphics[width=\textwidth]{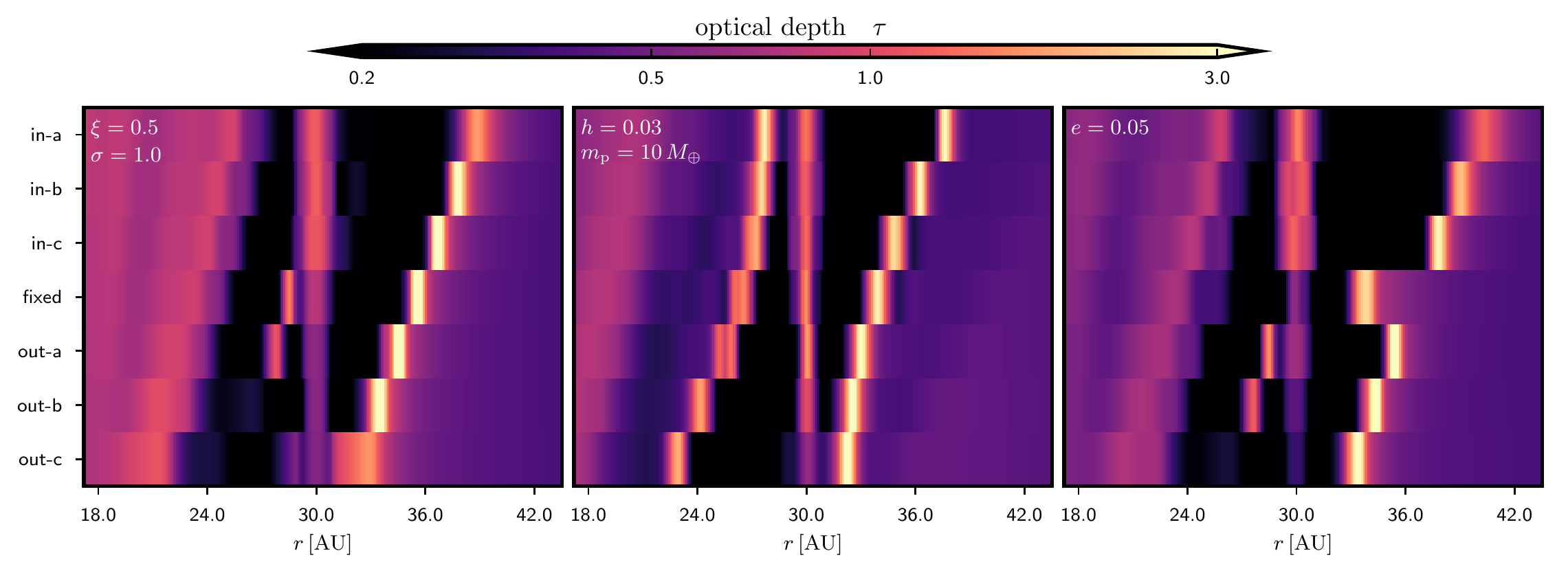}
	\caption{Azimuthal average of optical depth for a flared disk with $T\propto r^{-0.5}$ and $\Sigma_g\propto r^{-1}$ (left panel), a cold disk with $h=0.03$ (central panel) and an eccentric planet with $e=0.05$ (right panel). With the exception of the ones stated explicitly, all model parameters are those of the fiducial setup described in Table~\ref{tab:para}.}
	\label{fig:h-e}
\end{figure*}

In our disk model we assumed a power-law dependency of the disk temperature and density on the distance to the central star, $T \propto r^{\xi}$ and $\Sigma \propto r^{\sigma}$, with the fiducial power-law exponents of $\xi=-1.0$ and $\sigma=-0.5$. This implies that the gas disk's aspect ratio, $h$, is constant. To account for the case of a flared disk profile, with $h$ increasing with the distance to the star, we show the results for $\xi=-0.5$ and $\sigma=-1.0$ in the left panel of Figure~\ref{fig:h-e}. As the figure shows, the locations of gaps and rings is hardly affected by this choice. 

Reducing the aspect ratio of the PPD is equivalent to considering a colder disk. In such an environment perturbations are less efficiently damped and gap formation is more pronounced. To investigate the effect this has on the structure created by migrating planets we reduce the aspect ratio to $h=0.03$. \citet{2006Icar..181..587C} note that the pressure torque of the disk is proportional to $h^2$ and that therefore -- as long as the pressure torque is relevant for dynamical processes -- a smaller value of $h$ requires a steeper gap profile in an equilibrium state. In other words, this means that the disk is more sensitive to the planet's perturbation for the reason that its strength to counteract these perturbations is diminished. This has two consequences: first, already a quite small planet mass can produce observable structure.
We highlight this in the central panel of Figure~\ref{fig:h-e} for the case of an aspect ratio of $h=0.03$ and $m_\mathrm{p}=10M_\oplus$. Note, that with this reduced aspect ratio, the Hill radius of this planet $R_\mathrm{H} = r_\mathrm{p}(m_\mathrm{p}/(3m_\ast))^{1/3}$ is already larger than the disk pressure scale height, $H$. Again, the figure shows the optical depth for different migration rates. Aside from migration features, two things become generally visible: Each single ring at the outer and inner edge of the gap are more pronounced and more narrow than in the corresponding fiducial case of $h=0.05$. Secondly, the relative spacing itself of gaps and rings is also more narrow than before. This is in accordance with the predictions by \citet{2001ApJ...552..793G} for the dependency of the shock locations on the scale height of the disk. \\

The second consequence of a low aspect ratio is the increased tendency towards developing azimuthal structure. In the case of $m_\mathrm{p}=17M_\oplus$ and $h=0.03$ for example, the edges of the gap become steep enough to trigger Rossby-wave instabilities \citep{1999ApJ...513..805L,2000ApJ...533.1023L} in the gas phase that disrupt the azimuthal structure from being symmetric. The creation of a ringed structure is therefore limited by the disk's thermal mass, $m_\mathrm{p} \lesssim m_\mathrm{th}=m_\ast h^3$, and the proposed method to infer migration behavior from concentric ring structure breaks down if the planet exceeds this mass.  

\paragraph{Eccentricity}
Eccentricity both affects the dust structure in a PPD (as we will show), but it is also a parameter that changes the migration behavior itself \citep[e.g.][]{2000MNRAS.315..823P,2011ApJ...737...37M}.
For example, thermal excitation \citep{2017A&A...606A.114C,2017MNRAS.469..206E,2019MNRAS.485.5035F} or planet-planet interaction \citep[e.g.][]{2002ApJ...567..596L} can cause eccentric orbits. We simulate several migration scenarios, considering a planet with a fixed and prescribed eccentricity, $e$. We implicitly assume that the source of eccentricity excitation does not influence the dust structure itself in the region of interest.

The right panel of Figure~\ref{fig:h-e} shows the optical depth for $e=0.05$ . It becomes apparent that in the eccentric case the gap is less pronounced in depletion but typically wider. Prominently, in the non-migrating case the dust accumulates at the same locations for both eccentric and circular orbits. This changes once the planet is migrating. Here we show that for an eccentric orbit the outer ring is even further away from the planet's position than before.    
This has to be kept in mind when attempting to infer the migration rate from ring locations, as the eccentricity of the planet introduces a further level of degeneracy to that.

\paragraph{The equation of state}
In our models a prescribed, locally isothermal temperature structure was employed.
Only recently, \citet{2019ApJ...878L...9M} emphasised that abandonning the isothermal assumption can have severe consequences for the resulting dust density structure. They showed this by choosing an ideal-gas equation of state, $P=(\gamma-1)\epsilon\Sigma_\mathrm{g}$. Here, $\epsilon$ is the volumic internal energy and they set the adiabatic index to $\gamma=1.001$, which differs from the isothermal value of $\gamma=1$ only by a small factor. The main difference is that, in addition to the continuity and momentum equations, an energy equation has to be advanced in time that allows the temperature in the disk  to change. \cite{2019ApJ...878L...9M} report that
this treatment can already develop fundamental differences in the gas density profile of the PPD.

Motivated by these findings, we study the behavior of the gas and dust densities in response to a migrating planet for two values of the adiabatic index, $\gamma=1.001$ and $\gamma = 1.4$. Most importantly, this treatment requires to solve the temporal evolution of the volumic internal energy additionally:
\begin{equation}\label{eq:e-ad}
	\partial_t \epsilon +\nabla\cdot (\epsilon\mathbf{v})= -P \nabla\cdot\mathbf{v}\,. 
\end{equation}
For the adiabatic model, the speed of sound is initially the same as in the vertically isothermal case, $c_\mathrm{s}= h\Omega_\mathrm{K}r$.
This initializes the volumic internal energy as $\epsilon=c_\mathrm{s}^2\Sigma_\mathrm{g}/(\gamma-1)$.
How the temporal evolution given by Equation~(\ref{eq:e-ad}) is then treated numerically is described in \citet{2016ApJS..223...11B}.

\begin{figure}
	\centering
	\includegraphics[width=\columnwidth]{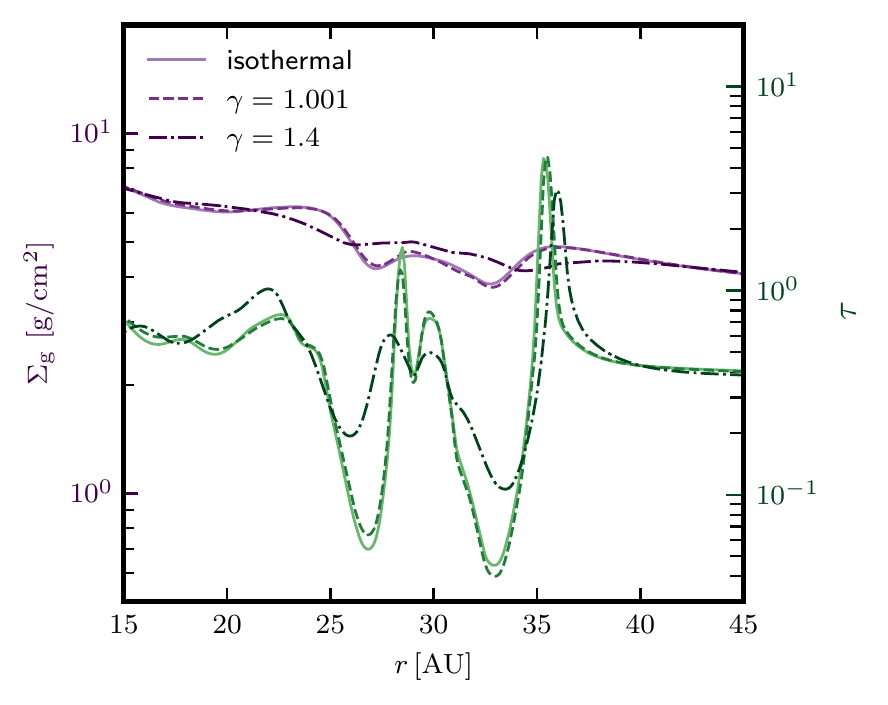}
	\caption{Gas surface density (left axis) and optical depth (right axis) for models with different equations of state. The solid lines show the vertically isothermal model, while the dashed and dash-dotted lines show the results of an adiabatic equation of state, with $\gamma=1.001$ and $\gamma=1.4$, respectively.}
	\label{fig:ad-comparison}
\end{figure}
Figure~\ref{fig:ad-comparison} shows how the radial gas and dust structures differ for the different treatments of the internal energy. The locally isothermal reference model is the fiducial setup, represented in Table~\ref{tab:para}, and for the adiabatic models the only thing that is changed is the equation of state.
As Figure~\ref{fig:ad-comparison} shows, the case of $\gamma=1.001$ hardly produces a difference to the isothermal model. Besides the relatively short evolution time of our system, one reason for the small change in this scenario might be that our focus is directed to the immediate vicinity of the planetary orbit, while \citet{2019ApJ...878L...9M}  report most significant effects of the EoS for $r<0.5 r_\mathrm{p}$. Additionally, the planetary mass in our simulations, corresponding to $m_p \approx 0.4 m_{\rm th}$, lies in an intermediate regime for which \citet{2019ApJ...878L...9M} find the effects of the EoS to not be as significant as for planet masses much smaller or larger than the thermal mass. For a value of $\gamma=1.4$, we find that the contrast between accumulation and gap gets significantly reduced. An additional effect is the widening of the gap, as both inner and outer dust rings are further away from the planet.
\begin{figure}
	\centering
	\includegraphics[width=\columnwidth]{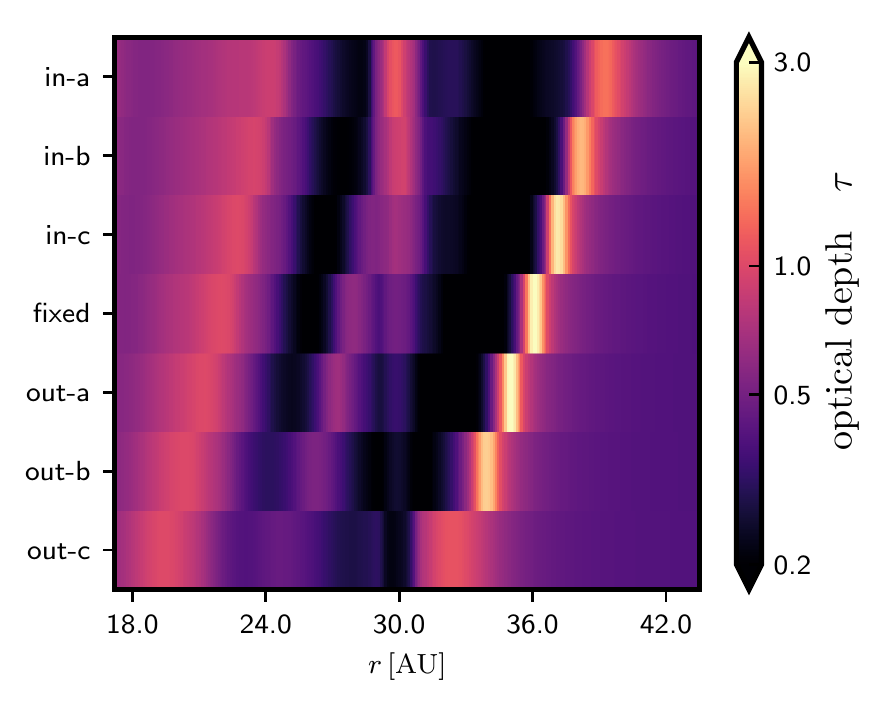}
	\caption{Same as Figure~\ref{fig:fid-kdens} but for an adiabatic equation of state with $\gamma=1.4$.}
	\label{fig:ad-kappadens}
\end{figure}
In Figure~\ref{fig:ad-kappadens} we show the migration - optical depth map for this case. For isolating the effect of the adiabatic treatment, this has to be directly compared to Figure~\ref{fig:fid-kdens}. In general the results show the same trend as described for 
Figure~\ref{fig:ad-comparison}: the structure is slightly wider and the contrast between rings and gaps is less pronounced. Nevertheless, the main features that a migrating planet invokes in the dust structure are unchanged. \citet{2019ApJ...878L...9M} mention that the decisive parameter for the correct treatment is the dimensionless cooling time $\Omega_\mathrm{K}t_\mathrm{c}$ compared to the disk's aspect ratio. As shown in \citet{2015ApJ...811...17L}, this parameter depends strongly on the distance to the star, which is why we expect the correct equation of state to be different depending on the region of the disk one wants to model. In general, the varying results highlight that for an improvement of presented studies, a better understanding of the equation of state is essential. 

\section{Discussion}\label{sec:discussion}
\paragraph{The effect of dust evolution and composition} We assumed in this work that solids are predominantly representable as compact, spherical dust grains in a PPD. The simulations do not account for dust evolution within the hydrodynamical models, i.e. there is no mass transfer between different grain sizes. The different possible outcomes for colliding dust grains is comprehensively summarised in \citet{2018SSRv..214...52B}. The grain growth is believed to be limited by both fragmentational collisions and so-called bouncing collisions (collisions without mass transfer).

It is also important to discuss the validity of the assumption that dust grains are compact in their nature. This is especially relevant in consideration of the idea that particles grow to fluffy aggregates \citep{2009ApJ...702.1490W,2012ApJ...752..106O,2013A&A...554A...4K}, which allows them to accumulate mass while still staying coupled to the gas dynamics. This mechanism relies on a high sticking behavior of dust grains, which are typically assumed to be icy at locations relevant to our work. Recently, \citet{2019ApJ...873...58M} investigated this property in the laboratory for mm-sized water ice grains and found  that their stickiness at very low temperatures is similar to that of silicates, thus unfavorable for fluffy grain growth. But whether this collisional behavior is also prevalent for sub-mm particles is still an open question.

The important quantity of dust grains to define their dynamics in a gaseous environment is the product of particle size, $a$, and intrinsic material density, $\rho_\mathrm{mat}$. In the results of hydrodynamical simulations that we presented, the material density was kept fixed at a value of $\rho_\mathrm{mat}=2\,\mathrm{g}\mathrm{cm}^{-3}$, the outcome of these simulations can be easily adapted to other values of this parameter by changig the assumed particle size adequately such that $a\times\rho_\mathrm{mat}$ stays unchanged.  The bouncing of grains also has an important significance in this framework: \citet{2009ApJ...696.2036W} showed from laboratory experiments, that although bouncing does not add or subtract mass from grains, it still leads to compactification of dust agglomerates, limiting the so-called filling factor to an equilibrium value of $f\approx 0.36$. This value can be incorporated into our hydrodynamical model by assuming an approximately three times smaller material density.

For the radiative transfer models, we assumed the dust grains to consist of silicates and amorphous carbon. It is valid to discuss the effect a change in the dust composition can have on the synthetic images. As stated in \citet{2019ApJ...877L..18Z} and \citet{2019ApJ...877L..22L}, an increased scattering opacity of dust grains (e.g. if they are covered in water ice) can have an important effect on the appearence of a disk. More precisely, if scattering is enhanced, the emission from optically thick regions of the disk is reduced. While potentially changing the observed fluxes and contrasts of presented predictions, the overall structure should remain unaffected by this.

\paragraph{Ring shape and location} In general it becomes apparent that the relative spacing of created dust accumulations depends strongly on the migration behavior of the planet. In all presented snapshots the planet is located at $30\,\mathrm{AU}$ so that the positions of the rings are directly comparable to each other. We find that the distance between the horseshoe ring and the ring in the direction the planet is migrating towards is diminished, while the distance towards the ring in the opposite direction is enhanced. The explanation for this is simple, the disk structure is inert to the updated planet position. In some extreme cases of fast outward migration the planet even approaches the outer ring close enough to destroy the pressure trap there and trigger the transport of dust towards the inner system.   

The origin of the inner ring is not necessarily due to a true pressure trap, i.e. at this location the pressure gradient may still be directed inwards and therefore the azimuthal gas velocity sub-Keplerian. As a matter of fact, this is the case in the fiducial model, displayed in Figure~\ref{fig:azvel}. The varying location of the inner ring for different particle sizes can be explored by multiple wavelength observations. If there is a true pressure trap at this location (i.e. a radial area in which the azimuthal gas velocity turns super-Keplerian) the different dust species are expected to pile up at the same location. This can be seen in the central panel of Figure~\ref{fig:h-e}, where for a reduced aspect ratio this criterion is fulfilled. Here, the observed ring should be traced to the same location independent of the wavelength of observation. In contrast, Figure~\ref{fig:nomig2d} predicts that increasing the wavelength for the fiducial model will shift the inner ring location to slightly smaller radii, due to the higher sensitivity for larger dust grains.

\paragraph{Application to observations}
The intent of this work is to learn from observations, both about the level of viscosity and about the migration behavior of proposed planetary candidates in PPDs. It stands to reason to start by looking at objects that already exhibit structures that suggest such planetary cores. Two famous examples are HL Tau and HD~169142, both featuring a narrow double gap structure. \citet{2018ApJ...866..110D} have already shown that a medium mass planet in a low-viscosity disk can produce a structure comparable to the double gap in HL Tau. The viscosity parameter estimate by \citet{2016ApJ...816...25P} ($\alpha\approx 10^{-4}$) for this system is in agreement with the idea that this structure is created by a single embedded planet. Considering the symmetric spacing of the gaps and rings, we can infer by comparison to our presented results that, if the structure is really due to a planet, the object cannot be migrating rapidly, neither inwards nor outwards. 

The case of HD~169142 has already been studied under the aspect of planetary migration by \citet{2019AJ....158...15P}. While we can confirm the result that an inward migrating planet creates asymmetric triple ring structure, we point to a certain degeneracy when inferring a discrete migration rate. Our results suggest that while the overall pattern remains intact, parameters such as aspect ratio and planetary eccentricity change the structure of dust accumulations, as well. Without precise knowledge of the systems parameters the exact migration rate is hard to isolate. Yet potential future detections and further characterisations of already known environments can help to reveal the most principal tendencies.

\section{Conclusions}
The presented work shows that different migration scenarios of an embedded planet can be associated with visible structure in highly resolved observations. In the parameter space in which a single planet causes circular, concentric dust accumulations in multiple locations in the disk, the precise spacing of such locations, both with respect to the planet and with respect to each other, can change for different migration scenarios. We qualitatively outlined, how this link can be used to put constraints on the migration direction and rate in PPDs. The method's big advantage is that it does not depend on relative fluxes of different rings and is therefore relatively robust against different thermal disk models. Future observations can help in shedding light on the long standing questions whether and how planets move radially during their formation and will help to promote or refute certain theoretical models of planet migration. 
\quad\\

\acknowledgements
We thank the anonymous referee for useful comments and suggestions that improved the quality of the presented work.
PW is grateful for the hospitality and the support experienced at Universidad de Chile and Universidad de Santiago de Chile, where the ideas leading to the presented work were developed. The research leading to these results has received funding from the European Research Council under the European Union's Horizon 2020 research and innovation programme under grant agreement No 638596 (PW, OG, LK). SP and SC acknowledge financial support from CONICYT-FONDECYT grant numbers 1171624 and 1191934.
This project has received funding from the European Union's Horizon 2020 research and innovation programme under grant agreement No 748544 (P.B.L.). 
The hydrodynamical simulations were performed on GPU computing nodes funded with a research grant from the Danish Center for Scientific Computing and are hosted at the University of Copenhagen HPC facility. This work used the Brelka cluster (FONDEQUIP project EQM140101) hosted at DAS/U. de Chile.

\software{
This work has made use of FARGO3D \citep{2016ApJS..223...11B,2019ApJS..241...25B} for hydrodynamical simulations, RADMC3D \citep{2012ascl.soft02015D} for radiative transfer calculations, 
the casa-5.5.0 software package \citep{2007ASPC..376..127M} for simulating synthetic observations
as well as 
IPython \citep{ipython}, NumPy \citep{numpy} and Matplotlib \citep{Matplotlib} for data analysis and creating figures.}

\appendix
\section{Numerical treatment of dust diffusion}\label{appendix}
\subsection{Implementation in FARGO3D}
To account for microscopical stochastic kicks from the turbulent gas onto the dust, we implement diffusion as a source term in the continuity equation shown in Equation~(\ref{eq:contdust}) following the suggestion in \citet{2019ApJS..241...25B}. We emphasize that this is pure mass diffusion, there is no additional macroscopic force added for the dust fluids and the macroscopic velocity $\mathbf{v}$ is left unaltered. Under this consideration, there is no additional term in the non-conservative form of the momentum equation as it is written in Equation~(\ref{eq:NS-dust}).
However, when treating the dust momentum equations in their conservative form an additional term arises to account for the shift of momentum due to the shift of mass. This can be shown by multiplying Equation~(\ref{eq:contdust}) with the dust velocity and adding it to Equation~(\ref{eq:NS-dust}), which results in:
\begin{equation}
	\partial_t(\rho\mathbf{v}) + \mathbf{\nabla}(\rho\mathbf{v}\cdot\mathbf{v}) = \mathbf{S} - \mathbf{v}\mathbf{\nabla}\cdot\mathbf{j}\,,\label{eq:momeqcon}
\end{equation}
where we replaced the surface density, $\Sigma_{{i}}$, by the more general expression, $\rho$, and the sources on the RHS of Equation~(\ref{eq:NS-dust}) are substituted by $\mathbf{S}$. The expression $-\mathbf{v}\mathbf{\nabla}\cdot\mathbf{j}$ is the contribution arising from mass diffusion. 

We choose to solve the diffusion step within the source step \citep[see][]{2016ApJS..223...11B}, by updating only the density using $\partial_t \rho_\mathrm{d} = -\mathbf{\nabla}\cdot \mathbf{j}_\mathrm{diff}$. This is done to first order explicitly. We do not require an additional update to account for the extra term in the momentum because the shift in momentum is already accounted for due to the operator splitting method used by the code.

\begin{figure}[h]
	\centering
	\includegraphics[width=0.33\columnwidth]{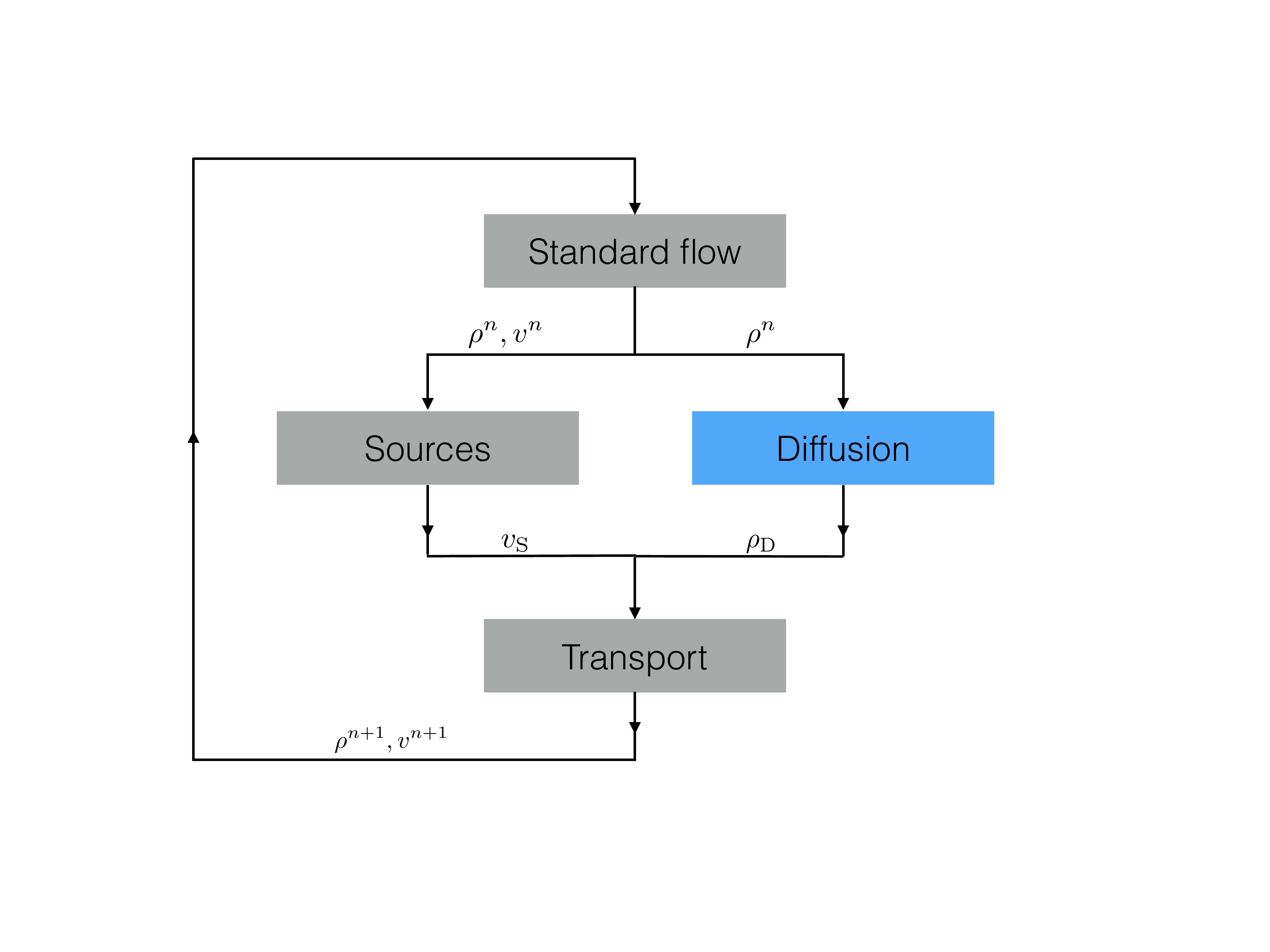}
	\caption{Flowchart of FARGO3D with dust diffusion. In praxis the diffusion update is applied after the source step.}
	\label{fig:flowchart}
\end{figure}
Figure~\ref{fig:flowchart} shows the order in which important updates are performed during one time step. In the following we construct analytical solutions to test this implementation.

\subsection{Test A: Diffusion in Cartesian coordinates}
The first test we design is one-dimensional diffusion in Cartesian coordinates. The only coordinate is $x$, the dust density is represented by $\rho$. Further we assume, that the background fields are static and constant in space, that there are no external forces acting on the dust and gas mixture, that the diffusion coefficient is a constant and we neglect the feedback imposed by the dust onto the gas. The matrix representation of the linearized equations reads: 
\begin{equation}
-i\omega\, \left( {\begin{array}{c}
	\delta \rho \\
	\delta v \\
	\end{array} } \right) =
\left( {\begin{array}{cc}
	ikv + Dk^2(1-\varepsilon) & ik\rho \\
	0 & ikv+\frac{1}{\tau_\mathrm{s}} \\
	\end{array} } \right) \left( {\begin{array}{c}
	\delta \rho \\
	\delta v \\
	\end{array} } \right)\,,
\end{equation}
where we assume the frictional stopping time, $\tau_{\rm s}$, to be constant, $\varepsilon = \rho/ \rho_{\rm tot}$, and $\delta \rho = \delta \rho_{0}\exp(i(kx + \omega t))$ and $\delta v = \delta v_0 \exp(i(kx+\omega t))$ are small perturbations to the background fields, with $\delta \rho_0 \ll \rho$ and $\de v_0 \ll v$.
For numerical tests we choose the background velocities $v_0 = 1.0$ and $u_0=1.0$ for gas and dust, respectively. The background dust density is $\rho_0=1.0$ and the total background density is $\rho_{\mathrm{tot},0}=2.0$. 
This matrix equation leads to two solutions for the dispersion relation.

\begin{figure*}
	\centering
	\includegraphics[width=\textwidth]{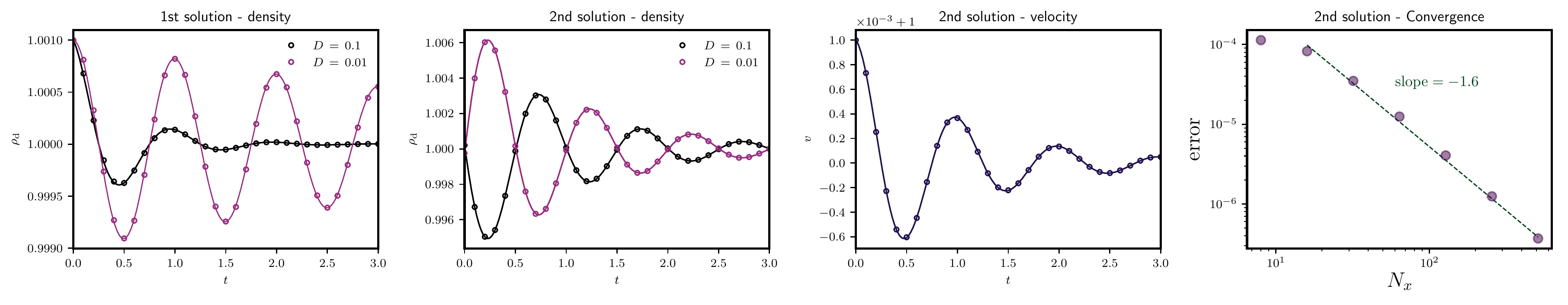}
	\caption{Linear tests of diffusion in Cartesian coordinates. Circles correspond to numerical solutions, solid lines to analytical solutions presented in Equation~\ref{eq:solution1} (first panel), Equation~\ref{eq:sol2dens} (second panel) and Equation~\ref{eq:sol2vel} (third panel) for two different diffusion coefficients, $D$, in each case. Note, that the solution for the velocity in Equation~\ref{eq:sol2vel} is independent of $D$. The panel on the right shows the error defined in Equation~\ref{eq:error} as a function of number of cells, $N_x$, for the case of the second solution with $D=0.1$.}
	\label{fig:cartesian}
\end{figure*}

\paragraph{First solution}
The trivial solution is setting the velocity perturbation to zero, $\de v = 0$.
The dispersion relation in this case is 
\begin{equation}
\omega_1 = -kv + iDk^2\left(1-\varepsilon\right)\,.
\end{equation} 
And the perturbed state is defined by:
\begin{eqnarray}\label{eq:solution1}
\dr \,\,&=&\,\, \de \rho_{0} \exp(-Dk^2\left(1-\varepsilon\right) t)\cos (kx \pm kvt)\,,\\
\de v \,\,&=&\,\, 0\,.
\end{eqnarray}
The left panel of Figure~\ref{fig:cartesian} shows both the numerical and the analytical solution of the evolution of the dust density  for $\delta \rho_0 = 0.001$ at the location $x=0$.

\paragraph{Second solution}
The second solution yields the dispersion relation
\begin{equation}
\omega_2 = -kv+\frac{i}{\tau_\mathrm{s}}\,,
\end{equation}
and the relation between the perturbations
\begin{equation}
\delta \rho = -\delta v \frac{ik\rho_\mathrm{0}}{Dk^2(1-\varepsilon)-\tau_\mathrm{s}^{-1}}\,.
\end{equation}
Solving these equations for the real parts of the perturbations gives us:
\begin{eqnarray}
\mathrm{Re}(\delta v)\;\;& =& \;\; \delta v_0 \cos(kx-k v t) e^{-t/\tau_\mathrm{s}}\,\label{eq:sol2dens}\\
\mathrm{Re}(\delta \rho) \;\; &=& \;\; \delta v_0 \frac{k\rho_\mathrm{0}}{Dk^2(1-\varepsilon)-\tau_\mathrm{s}^{-1}}\sin( kx- k v t ) e^{-t/\tau_\mathrm{s}}\,.\label{eq:sol2vel}
\end{eqnarray}
In Figure~\ref{fig:cartesian}, the two central panels show that the numerical solutions of density and velocity recover the analytical ones. The test was performed for $\delta v_0 = 0.001$ .
\paragraph{Convergence Test}
Finally we want to investigate the dependence of the error on the resolution. For this we define the error of the simulation as
\begin{equation}\label{eq:error}
	\mathrm{error} = \frac{1}{N_x}\sqrt{\sum_{i}^{N_x}\left\langle\frac{\rho_i-\hat{\rho}_i}{\hat{\rho}_i}\right\rangle^2}\,,
\end{equation}
where $N_x$ is the number of cells, and $\hat{\rho}_i$ is the analytical solution for the density in a specific cell, $i$. We measure the error after a time of $t=1.0$, fixing the time step to a value of $dt=10^{-6}$ to have the same time step for all the resolutions. We ensure that this time step is in all cases well beyond the CFL-condition. The result of this test is presented in the right panel of Figure~\ref{fig:cartesian}.

\subsection{Test B: Diffusive spreading in Cylindrical Coordinates}
To test the diffusion implementation in cylindrical coordinates we consider the spreading of a concentric ring of dust in a gaseous environment. The derivation is quite similar to \citet{1974MNRAS.168..603L} for viscous spreading. To isolate the effect of diffusion we assume that $\rho_\mathrm{tot}$ is approximately constant in time and space, which implicitly demands $\rho \ll \rhog$, since the evolution of $\rho$ is what we are interested in. Further we set $u=v=0$ and $D=\mathrm{constant}$. The continuity equation of the dust then reads:
\begin{equation}\label{equ:testb1}
	\partial_t \rho = \frac{1}{r} \partial_r \left(rD\rho_\mathrm{tot}\partial_r\left(\frac{\rho}{\rho_{\mathrm{tot}}}\right)\right) = \frac{D}{r} \partial_r \left(r\partial_r\left(\rho\right)\right)=D\partial^2_r\rho + \frac{D}{r}\partial_r \rho\,.
\end{equation}
We assume $\rho$ to be a function of time and space, $\rho = T(t)R(r)$. We divide Equation~\ref{equ:testb1} by $\rho$ and obtain:
\begin{equation}
	\frac{\partial_t T(t)}{T(t)} = \frac{D\partial_r^2R(r)}{R(r)} + \frac{D}{rR(r)}\partial_r R(r)\,.
\end{equation}  
We can separate this equation:
\begin{equation}
	\frac{\partial_t T(t)}{T(t)} = -k\,, \qquad r^2\partial_r^2R(r) + rD\partial_r R(r) + \frac{k}{D}r^2R(r) =0 \,,
\end{equation}
where $k$ is a constant.
The time dependent part gives us:
\begin{equation}
	T(t) = T_0 \exp(-kt)\,,
\end{equation}
which gives the constraint that $k>0$.
For the radial part we substitute $\tilde{r}\equiv \sqrt{k/D} r$ and recognize the Bessel differential equation:
\begin{equation}
	\tilde{r}^2\frac{\partial^2}{\partial \tilde{r}^2} R(\tilde{r}) + \tilde{r} \frac{\partial}{\partial \tilde{r}} R(\tilde{r}) + \tilde{r}^2 R(\tilde{r}) = 0 \,,
\end{equation}
which is solved by
\begin{equation}
	R(\tilde{r}) = A(k) J_0(\tilde{r}) + B(k) Y_0(\tilde{r})\,.
\end{equation}
where $J_0$ is a Bessel function of the first kind and $Y_0$ is a Bessel function of the second kind. We demand the density to be finite for $r=0$, which gives the constraint that $B(k)=0$. 
The general solution for $\rho$ is then:
\begin{equation}
	\rho(r,t)=\int_{0}^{\infty} \exp(-kt) A(k) J_0\left(\sqrt{\frac{k}{D}} r\right) dk \, .
\end{equation}
We then write $c=\sqrt{k/D}$ and obtain:
\begin{equation}\label{eq:general_solution}
	\rho(r,t)=\int_{0}^{\infty} \exp(-D c^2t) A(c) J_0(c r) 2 c D dc \, ,
\end{equation}
and recognize the Hankel-transform of order zero:
\begin{equation}
	\mathcal{H}_0(A(c)) = \int_{0}^{\infty} A(c)J_0(cr)c dc = \frac{\rho(r,t=0)}{2D}\,,
\end{equation}
for which the inverse transformation delivers the coefficient $A(c)$:
\begin{equation}\label{eq:hankinv}
	A(c) = \mathcal{H}_0^{-1} = \int_{0}^{\infty} \mathcal{H}_0(A(c))J_0(cr)r dr = \frac{1}{2D}\int_{0}^{\infty} \rho(r,t=0) J_0(cr)r dr\,.
\end{equation}
We choose the most convenient initial distribution, $\rho_\mathrm{d}(r,t=0) = m\delta(r-r_0)$, where $m$ is an arbitrary normalization factor, and obtain $A(c)=m(2D)^{-1}J_0(cr_0)r_0$ by performing the integral in Equation~\ref{eq:hankinv}. This inserted into Equation~(\ref{eq:general_solution}) reads:
\begin{equation}
	\rho(r,t)= r_0m\int_{0}^{\infty}\exp(-Dc^2t)J_0(cr_0)J_0(cr)cdc = r_0^{1/2}m\int_{0}^{\infty}f(c)J_0(cr_0)(cr_0)^{1/2}dc\,,
\end{equation}
where we defined $f(c)=\exp(-Dc^2t)J_0(cr)c^{1/2}$. We can solve the latter integral by using a lookup table\footnote{\href{https://authors.library.caltech.edu/43489/7/Volume\%202.pdf}{Tables of Integrals}} (p. 51, Eq. (23)):
\begin{equation}\label{eq:radspread}
\rho(r,t) = \frac{mr_0}{2Dt} \exp\left(-\frac{r^2+r_0^2}{4Dt}\right) I_0\left(\frac{r r_0}{2Dt}\right)\,,	
\end{equation}
where $I_0$ is the modified Bessel function of the first kind.
We can use the result of Equation~(\ref{eq:radspread}) to test the diffusion implementation in radial direction. To not include an infinitesimally confined distribution we choose $\rho(r,t=2)$ as initial condition and evolve the density in time. We set $D=10^{-3}$, $m=2\times10^{-6}$, $r_0=1.0$, $N_r=1000$ and $r=[0.1,2.5]$. 
\begin{figure*}
	\centering
	\includegraphics[width=\textwidth]{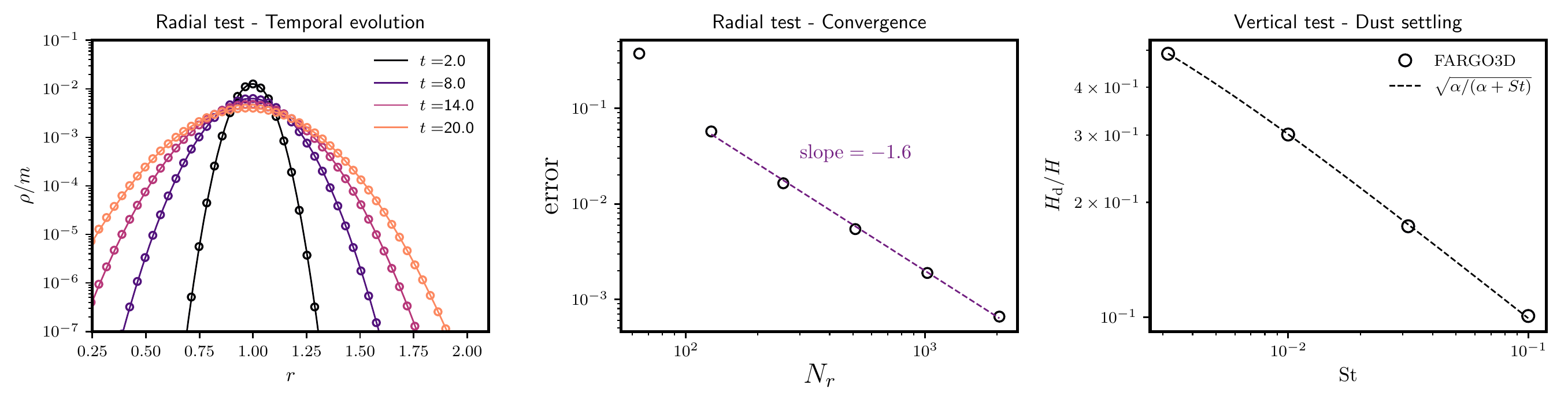}
	\caption{The left panel shows the temporal evolution of dust density in cylindrical coordinates. The circles correspond to the numerically calculated values, the solid lines are the analytical solutions presented in Equation~(\ref{eq:radspread}). In the central panel we test for convergence, plotting the error as defined in Equation~(\ref{eq:error}) depending on the number of radial cells. In the right panel we show the ratio of dust- and gas scale height depending on the Stokes number of the dust species. Again, we show the numerical measurement in circles and the analytical formula as a dashed line.}
	\label{fig:radspread}
\end{figure*}
The left panel of Figure~\ref{fig:radspread} compares numerical (circles) and analytical (lines) solutions at different times, the black line showing the initial condition. Finally, we show in the central paner of Figure~\ref{fig:radspread} that the error decreases for higher resolutions and the numerical solution converges to the analytical one.

\subsection{Test C: Dust settling in spherical coordinates}
To perform a test applied to the vertical profile of a PPD in spherical coordinates, we set up a simulation considering the vertical profile in hydrodynamical equilibrium. In the local isothermal case, the vertical structure of the gas follows a Gaussian profile, characterized by the pressure scale height, $H$:
\begin{equation}
	\rho_\mathrm{g} (\theta)= \rho_\mathrm{g}\left(\theta=\frac{\pi}{2}\right)\times\exp\left(-\frac{r^2\cos(\theta)^2}{2H^2}\right)\,.
\end{equation}
The vertical profile of the dust in steady-state is expected to equally follow a Gaussian, with the equivalent function:
\begin{equation}
\rho_\mathrm{d} = \rho_\mathrm{d}\left(\theta=\frac{\pi}{2}\right)\times\exp\left(-\frac{r^2\cos(\theta)^2}{2H_\mathrm{d}^2}\right)\,,
\end{equation}
where it is possible to derive, that $H_\mathrm{d}/H=\sqrt{\alpha/(\alpha+\mathrm{St})}$ \citep[see][]{1995Icar..114..237D}. 
We set up a background gas disk in hydrodynamicacl equilibrium and initialize the dust density distribution with the same profile, but reduced by a factor of $\varepsilon=0.01$. We choose a non-constant diffusion coefficient $D=\alpha c_\mathrm{s}^2/\OmegaK$, with $\alpha=10^{-3}$, and plot the ratio $H_\mathrm{d}/H$ at the location $r=1.0$ for different Stokes numbers in Figure~\ref{fig:radspread}. The numerical values are measured after a time of $t = \tau/\mathrm{St}$, with $\tau = 10\, \mathrm{orbits}$, to ensure that the dust has settled to an equilibrium at the time of measurement. The figure shows both the numerical and analytical solutions.
\end{document}